\documentclass[11pt]{article}
\pdfoutput=1

\usepackage{aas_macros,amsmath,amssymb,comment,cite,esint,graphicx,mathtools}
\usepackage[margin=.8in,letterpaper]{geometry}
\usepackage[colorlinks=true]{hyperref}
\usepackage[affil-it]{authblk}
\usepackage{url}
\usepackage{ulem}

\numberwithin{equation}{section}
\let\originalleft\left
\let\originalright\right
\renewcommand{\left}{\mathopen{}\mathclose\bgroup\originalleft}
\renewcommand{\right}{\aftergroup\egroup\originalright}
\mathcode`\*="8000
{\catcode`\*=\active\gdef*{\mathclose{}\,\mathopen{}}}

\newcommand{\ab}[1]{\left|#1\right|}
\newcommand{\br}[1]{\left[#1\right]}
\newcommand{\cu}[1]{\left\{#1\right\}}
\newcommand{\pa}[1]{\left(#1\right)}
\newcommand{\td}[1]{\tilde{#1}}
\newcommand{\M}[1]{\mathcal{#1}}

\newcommand{\be}{\begin{equation}}
\newcommand{\ee}{\end{equation}}
\newcommand{\bea}{\setlength\arraycolsep{2pt} \begin{eqnarray}}
\newcommand{\eea}{\end{eqnarray}}

\def \a {\alpha}
\def \b {\beta}
\def \g {\gamma}

\def \G {\Gamma}

\def \D {\Delta}
\def \e {\epsilon}
\def \ve {\varepsilon}
\def \m {\mu}
\def \n {\nu}
\def \k {\kappa}
\def \l {\lambda}
\def \L {\Lambda}
\def \s {\sigma}
\def \S {\Sigma}
\def \r {\rho}
\def \o {\omega}

\def \th {\theta}
\def \Th {\Theta}
\def \t {\tau}
\def \z {\zeta}

\def \p {\partial}
\def \f {\frac}

\begin{document}
\title{Photon emissions from near-horizon extremal and near-extremal Kerr equatorial emitters}

\author{
Haopeng Yan$^{1}$, Zezhou Hu$^{2}$, Minyong Guo$^{3\ast}$,
Bin Chen$^{1,2, 4}$}
\date{}

\maketitle

\vspace{-10mm}

\begin{center}
{\it
$^1$Center for High Energy Physics, Peking University,
No.5 Yiheyuan Rd, Beijing 100871, People's Republic of China\\\vspace{4mm}

$^2$Department of Physics, Peking University, No.5 Yiheyuan Rd, Beijing
100871, People's Republic of China\\\vspace{4mm}

$^3$ Department of Physics, Beijing Normal University,
Beijing 100875, People's Republic of China\\\vspace{4mm}

$^4$ Collaborative Innovation Center of Quantum Matter,
No.5 Yiheyuan Rd, Beijing 100871, People's Republic of China\\\vspace{2mm}
}
\end{center}

\vspace{8mm}

\begin{abstract}
We consider isotropic and monochromatic photon emissions from equatorial emitters  moving along future-directed timelike geodesics in the near-horizon extremal Kerr (NHEK) and near-horizon near-extremal Kerr (near-NHEK) regions, to asymptotic infinity.
We obtain numerical results for the photon escaping probability (PEP) and derive analytical expressions for the maximum observable blueshift (MOB) of the escaping photons, both depending on the emission radius and the emitter's proper motion. In particular,
we find that for all anti-plunging  or deflecting emitters that can eventually reach to asymptotic infinity, the PEP is greater than $50\%$ while for all plunging emitters the PEP is less than $55\%$, and for the bounded emitters in the (near-)NHEK region, the PEP is always less than $59\%$.
In addition, for the emitters on unstable circular orbits in the near-NHEK region, the PEP decreases from $55\%$ to $50\%$ as the orbital radius decreases from the one of the innermost stable circular orbit to the one of the horizon.
Furthermore, we show  how the orientation of the emitter's motion along the radial or azimuthal direction affects the PEP and the MOB of the emitted photons.
\end{abstract}

\vfill{\footnotesize Email:haopeng.yan,\,z.z.hu,\,
minyongguo,\,bchen01@pku.edu.cn.\\$~~~~~~\ast$ Corresponding author.}

\maketitle

\section{Introduction}\label{sec:Introduction}
Since the release of  the first image of the supermassive black hole M87*  by the Event Horizon Telescope (EHT) collaboration \cite{Akiyama:2019cqa} in 2019, there have been ongoing progresses in the community to explore physical information of the  black hole. This M87* image matches very well with the simulation of a rotating Kerr black hole based on the Einstein's theory of General Relativity.
Very recently, an improved image with polarization information was released, which would help us to explore the structure of the magnetic fields near the black hole \cite{EventHorizonTelescope:2021srq}.
Understanding these observations relies not only on the study of photon emissions and propagations in the black hole backgrounds, but also on the rich physics in the surroundings of the black holes,  such as the magnetohydrodynamics in plasmas, the distribution of magnetic fields and even the nature of dark matter halo. Consequently the images of the black holes may allow us to investigate various problems in the strong gravitational field region.

Besides the silhouette of a black hole shadow, there are other possible observational signatures with interesting features, for examples, the image of an hot spot surrounding a black hole \cite{Gralla:2017ufe,Guo:2018kis,Yan:2019etp,Guo:2019lur}, light rings (LRs) \cite{Cunha:2020azh,Guo:2020qwk,Cardoso:2021sip} and photon rings \cite{Gralla:2019xty,Himwich:2020msm,Johnson:2019ljv,Gralla:2020srx,Peng:2020wun,Peng:2021osd}.  Theoretically, as the first step, one has to understand clearly the photon emissions around the black hole, especially near the black hole horizon. On the other hand, studying the photon emissions from the near-horizon area is important to explore various high-energy processes\cite{Penrose:1969pc,Piran1975, Berti:2014lva,Schnittman:2014zsa,Zhang:2020tfz, Banados:2009pr, Guo:2016vbt,Zhang:2016btg,Blandford:1977ds,Camilloni:2020qah,Callebaut:2020hwv} happened in the near-horizon region (inside the ergosphere).

Recently, the photon escaping probability (PEP) and maximum observable blueshift (MOB)\footnote{We use the term ``MOB" for the reason that the photons with maximum net blueshift can always reach at infinity and thus are observable in principle. But in practice these escaping photons may not be observed by a distant observer since the MOB could be negative and the PEP could also be vanishingly small.} of photon emissions from near-horizon sources around rotating black holes had been studied by several groups. The PEP of a zero-angular momentum source (ZAMS) near a Kerr-Newmann black hole was first studied in \cite{Ogasawara:2019mir}, and later the study was generalized to the Kerr-Sen spacetime in \cite{Zhang:2020pay}. The PEP of a ZAMS approaching the event horizon is zero for a non-extremal Kerr black hole and tends to $\sim29\%$ for an extremal Kerr black hole.
Moreover, the PEP and MOB of photon emissions from equatorial (stable) circular orbiters near a Kerr black hole was studied in \cite{Igata:2019hkz} and \cite{Gates:2020els,Gates:2020sdh} with different methods clarifying the photon escaping condition.
The PEP for a circular orbiters increases as the orbital radius increases from  the innermost stable circular orbit (ISCO), while it decreases as the black hole spin parameter  increases at a given orbital radius. Remarkably, the value of PEP for an emitter sitting at the ISCO of an extremal Kerr black hole is still about $55\%$, and the proper motion of the emitter blueshifts the energy of the escaped photons.
Very recently, photon emissions from the plunging orbiters\footnote{The appearance of a bright hot spot falling into a non-spinning black hole and its relation with LRs were studied in Ref.~\cite{Cardoso:2021sip}.} starting from the ISCO of a black hole with arbitrary spin was  studied in \cite{Igata:2021njn}, and it was found that the PEP is always more than $50\%$ for the emitter at approximately halfway between the ISCO and the event horizon. Moreover, the effects of the proper motion of a plunging emitter on the PEP and the MOB were discussed in \cite{Igata:2021njn}.
While these studies applied to the  emitter of arbitrary radius and the black hole of arbitrary spin,
special attentions had been paid to the near-horizon and near-extremal cases in \cite{Gates:2020els} since such cases can be dealt with analytically.
In a previous paper \cite{Yan:2021yuo}, we reinvestigated the photon emissions from a ZAMS near a Kerr black hole by adapting the method proposed in \cite{Gates:2020els} and explored the observability of the very deep region in the near-horizon throat of a high-spin black hole. To that end, we analytically computed the PEP and the MOB of the ZAMSs both in the near-horizon extremal Kerr (NHEK) and the near-horizon near-extremal Kerr (near-NHEK) geometries and found that the PEP tends to $\sim13\%$ at the radius of the innermost photon shell.

In this paper, we would like to study the effects of the emitter's proper motion on the PEP and the MOB of the escaping photons, with the ZAMS as a ``static" reference \cite{Ogasawara:2019mir,Yan:2021yuo}. We will focus on the equatorial emitters that follow timelike geodesics in the (near-)NHEK\footnote{We use ``(near-)NHEK" for both NHEK and near-NHEK.} geometries. We will consider not only the marginally plunging emitters starting from the ISCO \cite{Igata:2021njn} but also other possible plunging emitters. Moreover, we will consider the emitters with other geodesic motions as well, including the anti-plunging, the deflecting and the bounded motions.
In \cite{Igata:2021njn}, only the marginally plunging emitters which has the critical angular momentum $l_s=l_{\text{ISCO}}$ and critical energy $\o_s=\o_{\text{ISCO}}$ had been studied, while the black holes were allowed to have arbitrary spin. Here, instead, we will only focus on the high-spin  black hole but consider the emitters with more possible values of $l_s$ and $\o_s$. Note that in the high-spin case the energy $\o_s$ of the emitter is constrained to be near the superradiant bound $\o_\ast=l_s/2M$ with the corrections appearing at order $\mathcal{O}(\e^q)$ [see Eqs.~\eqref{extremalCond} and \eqref{BHcoordinates}].
The timelike geodesics in the (near-)NHEK regions were well classified in \cite{Kapec:2019hro,Compere:2017hsi,Compere:2020eat,Compere:2021bkk} based on the conserved quantities \eqref{conservedNK} in the (near-)NHEK geometries.
As we restrict the emitters on the equatorial plane, the Carter constant for these emitters is set to zero, $Q_s=0$, thus each orbit of an emitter is specifically labeled by its (near-)NHEK energy $E_s$, angular momentum $L_s$ and radial orientation $s_r$.
We will consider the behaviors of the PEP and the MOB of the escaping photons emitted from the orbit  labeled by $(L_s,E_s,s_r)$ and study how they are affected when each of these parameters varies. Our main results are summarized in Eqs.~\eqref{zmobout}, \eqref{zmobin}, \eqref{zmobinside} and in Fig.~\ref{fig:escapeandmob}. In addition, we provide detailed illustrations and discussions for the results in Figs.~\ref{fig:escapetogether}, \ref{fig:asy1}, \ref{fig:asy2} and \ref{fig:asy3}.

The remaining part of this paper is organized as follows. In Sec.~\ref{sec:Geodesics}, we briefly introduce  the Kerr geometry and the geodesics in it, as well as its near-horizon and (near-)extremal limits. In Sec.~\ref{sec:setup}, we define the problem of photon emissions from an arbitrary near-horizon equatorial emitter, and find the equations to read the PEP and the MOB. In Sec.~\ref{sec:NKemitters}, we perform the computations for the (near-)NHEK emitters. In Secs.~\ref{sec:dependenceonR} and \ref{motioneffects}, we display our main results with plots and discuss them in detail. In Sec.~\ref{sec:summary}, we summarize and conclude this work.

\section{Geometry and geodesics}\label{sec:Geodesics}
In this section, we review the timelike and the null geodesics in the Kerr, NHEK and near-NHEK geometries \cite{Bardeen:1973tla,Kapec:2019hro,Compere:2017hsi,Compere:2020eat,Compere:2021bkk}. In particular, we review the classifications of the (near-)NHEK geodesics based on their constants of motions. The timelike geodesics provide the orbits for the photon emitters with various motions. For the null geodesics, we focus on the unstable spherical photon orbits, which correspond to the threshold between the escaping photons and the captured ones.

\subsection{Kerr geometry and geodesics}\label{subsec:Kerrgeodesics}
The Kerr metric in the Boyer-Lindquist coordinates $(t,r,\th,\phi)$ is given by
\be
\label{metric1}
ds^2=-\frac{\Sigma\Delta}{\Xi}dt^2+\frac{\Sigma}{\Delta}dr^2+\Sigma d\theta^2+\frac{\Xi\sin^2\theta}{\Sigma}\pa{d\phi-\frac{2Mar}{\Xi} dt}^2,
\ee
where
\be
\label{metric2}
\Delta=r^2-2Mr+a^2,\qquad
\Sigma=r^2+a^2\cos^2\theta,\qquad
\Xi=(r^2+a^2)^2-a^2\Delta\sin^2\theta.
\ee
This metric describes a rotating black hole with a mass $M$ and a spin parameter $a\equiv J/M$.
The event horizons of the black hole are located at
$r_\pm=M\pm\sqrt{M^2-a^2}$.

A free particle possesses four conserved quantities of motion: the mass $\mu$, the energy $\o$, the axial angular momentum $l$ and the Carter constant $Q$.
Using the Hamilton-Jacobi method, one can derive the four-momentum $p^\mu$ of this particle \cite{Bardeen:1973tla}
\bea
\label{kerrPr}
p^r&=&\f{1}{\S}\pm_r\sqrt{\M{R}(r)},\\
\label{kerrPth}
p^\th&=&\f{1}{\S}\pm_\th\sqrt{\Th(\th)},\\
\label{kerrPphi}
p^\phi&=&\f{1}{\S}\br{\f{a}{\D}[\o(r^2+a^2)-al]+\f{l}{\sin^2\th}-a\o},\\
\label{kerrPt}
p^t&=&\f{1}{\S}\br{\f{r^2+a^2}{\D}[\o(r^2+a^2)-al]+a(l-a\o\sin^2\th)},
\eea
where
\bea
\label{kerrRPotential}
\M{R}(r)&=&[\o(r^2+a^2)-al]^2-\D[Q+(l-a\o)^2+\mu^2r^2],\\
\label{kerrThpotential}
\Th(\th)&=&Q+a^2(\o^2-\mu^2)\cos^2\th-l^2\cot^2\th,
\eea
are the radial and angular potentials, respectively, and $\pm_r$ and $\pm_\th$ denote the radial and polar orientations, respectively.

For a massive particle, we have $\mu>0$ and $p^\mu=\mu\f{\p x^\mu}{\p \t}$ with $\t$ being the proper time.
For a massless particle (photon), we have $\mu=0$ and $p^\mu=\f{\p x^\mu}{\p\tau}$ with $\t$ being an affine parameter. For null geodesics, the energy $\o$ may be scaled out from the expressions of $p^\mu$ under a reparameterization and it is convenient to introduce a pair of impact parameters for the null geodesics:
\be
\label{impactparameters}
\l=\f{l}{\o},\qquad
\eta=\f{Q}{\o^2}.
\ee

\subsection{(Near-)NHEK geometry and geodesics}\label{subsec:NKgeodesics}
In order to study the near-horizon geometry of a (near-)extremal Kerr black hole, it is convenient to consider the near-extremal limit \cite{Kapec:2019hro,Hadar:2014dpa,Gralla:2015rpa}
\be
\label{extremalCond}
a=M\sqrt{1-(\e\kappa)^2},\qquad
0<\e\ll1,\qquad
\kappa\,\,\,\text{finite},
\ee
and work with the Bardeen-Horowitz coordinates \cite{Bardeen:1999px,Kapec:2019hro,Hadar:2014dpa,Gralla:2015rpa}
\be
\label{BHcoordinates}
T=\e^q\f{t}{2M},\qquad
R=\f{r-M}{\e^q M},\qquad
\Phi=\phi-\f{t}{2M},\qquad
0<q\leq1.
\ee

Under the transformations \eqref{BHcoordinates} and
taking the limit $\e\rightarrow0$ for $0<q<1$, the Kerr metric \eqref{metric1} yields the NHEK metric \cite{Bardeen:1999px}
\bea
\label{NKmetric}
ds^2&=&2M^2\Gamma\br{-R^2 dT^2+\frac{dR^2}{R^2}+d\theta^2+\Lambda^2(d\Phi+RdT)^2},\\
\label{GammaLambda}
&&\Gamma(\theta)=\f{1+\cos^2\theta}{2},\hspace{3ex} \Lambda(\theta)=\f{2\sin\theta}{1+\cos^2\theta}.
\eea
In the NHEK geometry, the event horizon is mapped to $R=0$ and the ISCO is mapped to $R=2^{1/3}$. However, since the NHEK metric is invariant under a dilation $(R, T)\rightarrow (\ve R, T/\ve)$ for an arbitrary constant $\ve$, the ISCO can actually stay at anywhere in the intrinsic NHEK geometry and  $R=2^{1/3}$ gets meaningful only after the NHEK region being glued onto the asymptotic far region \cite{Compere:2017hsi,Gralla:2015rpa}.

Under the transformations \eqref{BHcoordinates} and
taking the limit $\e\rightarrow0$ for $q=1$, we find that the Kerr metric \eqref{metric1} yields the near-NHEK metric\footnote{We will use the same coordinates $(T,R,\th,\Phi)$ [Eq.~\eqref{BHcoordinates}] to describe the near-NHEK geometry which has already been used for the NHEK metric \eqref{NKmetric}. In the following, we will also use the same notations for the conserved quantities $(E,L,C,\l,\eta)$  both in the NHEK and near-NHEK geometries.
However, these should be distinguishable from the context.} \cite{Bredberg:2009pv,Gralla:2015rpa}
\be
\label{nNKmetric}
ds^2=2M^2\Gamma\br{-(R^2+\k^2) dT^2+\frac{dR^2}{(R^2+\k^2)}+d\theta^2+\Lambda^2(d\Phi+RdT)^2}.
\ee
In the near-NHEK geometry, the event horizon is mapped to $R=\k$ and the ISCO is mapped to $R=\infty$ \cite{Compere:2017hsi,Gralla:2015rpa}.

Note that the NHEK metric \eqref{NKmetric} is equivalent to the near-NHEK metric \eqref{nNKmetric} with zero near-horizon temperature, $\k=0$ \cite{Bredberg:2009pv,Kapec:2019hro}. Therefore, in the following we will work with Eq.~\eqref{nNKmetric} for both the NHEK case ($0<q<1$ and $\k=0$) and near-NHEK case ($q=1$ and $\k\neq0$), for simplicity.

The four-momentum $p^\mu$ of a free particle in the (near-)NHEK limit becomes \cite{Kapec:2019hro,Compere:2020eat}
\begin{subequations}
\label{nnkgeodesics}
\bea
\label{nnkPr}
p^R&=&\pm_r\frac{\sqrt{\M{R}_n(R)}}{2M^2\G},\\
\label{nnkPth}
p^\th&=&\pm_\th\frac{\sqrt{\Th_n(\th)}}{2M^2\G},\\
\label{nnkPphi}
p^\Phi&=&\frac{1}{2M^2\G}\pa{-\f{R(E+LR)}{R^2-\k^2}+\f{L}{\L^2}},\\
\label{nnkPt}
p^T&=&\frac{1}{2M^2\G}\pa{\f{E+LR}{R^2-\k^2}},
\eea
\end{subequations}
where
\bea
\label{nnkRpotential}
\M{R}_n(R)&=&-CR^2+2ELR+E^2+(C+L^2)\k^2, \\
\label{nkThpotential}
\Th_n(\th)&=&C+\pa{1-\f{1}{\L^2}}L^2-2M^2\G\mu^2,
\eea
are respectively the radial and angular potentials in the NHEK region,
with $C$, $E$ and $L$ being the (near-)NHEK conserved quantities: the Casimir, the energy and the angular momentum, respectively.
The (near-)NHEK conserved quantities are related to the Kerr conserved quantities by \cite{Kapec:2019hro,Compere:2020eat}
\be
\label{conservedNK}
l=L,\qquad
\o=\frac{L}{2M}(1+\e^{q}\f{E}{L}),\qquad
Q=C+\frac{3}{4}L^2-\mu^2M^2.
\ee
For the null geodesics we have $\mu=0$, then we may define a pair of shifted impact parameters
\be
\label{shiftedImpacts}
\lambda_0=\frac{E}{L},\qquad
\eta_0=\f{C}{L^2},
\ee
which are related to the Kerr impact parameters \eqref{impactparameters} by\footnote{This means that the photon emissions are near the superradiant bound. Similar parameters have also been introduced in \cite{Gralla:2017ufe}.}
\be
\label{nkexpandimpacts}
\l= 2M(1-\e^{q}\l_0)\qquad
\eta=M^2(3+4\eta_0).
\ee

It is convenient to introduce a critical angular momentum \cite{Compere:2017hsi,Compere:2020eat}
\be
\label{Lstar}
L^\ast=l^\ast=\f{2}{\sqrt{3}}\sqrt{M^2\m^2+Q},
\ee
which is for the geodesic at the innermost stable spherical orbit (ISSO). In terms of this critical angular momentum, the Casimir $C$ can be rewritten as
\be
\label{Cs}
C=-\f{3}{4}[L^2-(L^\ast)^2].
\ee
We will restrict to the future-directed geodesics which require $p^T>0$ [{Eq.~\eqref{nnkPt}}], i.e.,
\be
\label{nkfuturedirected}
E+L R>0.
\ee

Next, by analyzing the root structures of the radial potential \eqref{nnkRpotential}, the equatorial orbits are classified in the  phase space $(L,E)$ regarding to their radial motions \cite{Kapec:2019hro,Compere:2020eat}. The classifications\footnote{Note that there was a forgotten region for the near-NHEK deflecting orbits in \cite{Compere:2020eat}, which was corrected in \cite{Compere:2021bkk}. Note also that we use ``anti-plunging" for the ``outward" orbits in \cite{Compere:2020eat}.} for the NHEK orbits and for the near-NHEK orbits are collected in Table~\ref{table:NKorbits} and Table~\ref{table:nNKorbits}, respectively, where $R_\pm$ and $R_0$ are the root(s) for the corresponding cases
\be
\label{nnkroots}
R_\pm=\f{E}{C}L\pm\f{1}{|C|}\sqrt{(L^2+C)(E^2+\k^2C)},\qquad
R_0=-\f{E^2+\k^2L^2}{2E L}.
\ee
The meaning of these distinct orbits are as follows \cite{Compere:2017hsi,Compere:2020eat}.
Each ISSO orbits at a constant radius. A plunging orbit enters the NHEK geometry from the (near-)NHEK boundary and eventually crosses the horizon in finite affine time. An anti-plunging orbit emerges out of the horizon and eventually leaves the (near-)NHEK boundary in finite affine time. A deflecting orbit falls in from the (near-)NHEK boundary, bounces off at the radial turning point, and returns to the boundary in finite affine time. Marginal (plunging/anti-plunging/deflecting) orbits exist only in the NHEK geometry and leave the NHEK geometry in infinite affine time. A bounded orbit stays in the (near-)NHEK region between the horizon and the radial turning point. The anti-plunging and deflecting orbits can escape from the (near-)NHEK to infinity while the plunging and the bounded ones can not. Note that all near-NHEK spherical orbits are unstable while all NHEK spherical orbits are ISSO \cite{Compere:2020eat}.
\begin{table}[h]
\centering
\caption{Future-directed geodesics in NHEK \cite{Kapec:2019hro,Compere:2020eat}.}
  \label{table:NKorbits}
\begin{tabular}{c c c c c }
  \hline  \hline
  Classification& $L$ & $E$ & $\pm_r$ & Range \\
  \hline
  ISSO& $L=L^\ast$ & $E=0$ & $ \pm1$ & $0\leq R\leq\infty$  \\
  \hline
  Marginal& $L>L^\ast$ & $E=0$ & $ \pm1$ & $0\leq R\leq\infty$  \\
  \hline
  Plunging& $L\geq L^\ast$ & $E>0$ & $ -1$ & $0\leq R\leq\infty$  \\
  \hline
  anti-plunging& $L\geq L^\ast$ & $E>0$ & $ +1$ & $0\leq R\leq\infty$  \\
  \hline
  Deflecting& $L> L^\ast$ & $E<0$ & $ \pm1$ & $R_+\leq R\leq\infty$  \\
  \hline
        & $-L^\ast<L<L^\ast$ & $E>0$ & $ \pm1$ & $0\leq R\leq R_+$  \\
  Bounded& $ L=-L^\ast$ & $E>0$ & $ \pm1$ & $0\leq R\leq R_0$  \\
        & $ L<-L^\ast$ & $E>0$ & $ \pm1$ & $0\leq R\leq R_-$  \\
  \hline  \hline
\end{tabular}
\end{table}
\begin{table}[h]
\centering
\caption{Future-directed geodesics in near-NHEK \cite{Kapec:2019hro,Compere:2020eat}.}
  \label{table:nNKorbits}
\begin{tabular}{c c c c c }
  \hline  \hline
  Classification& $L$ & $E$ & $\pm_r$ & Range \\
  \hline
  Spherical& $L>L^\ast$ & $E=-\k\sqrt{-C}$ & $ \pm1$ & $R=\f{\k L}{\sqrt{-C}}$  \\
  \hline
  Plunging& $L=L^\ast$ & $E\geq0$ & $ -1$ & $\k\leq R\leq\infty$  \\
          & $L> L^\ast$ & $E>-\k\sqrt{-C}$ & $ -1$ & $\k\leq R\leq\infty$ \\
  \hline
  anti-plunging& $L=L^\ast$ & $E\geq0$ & $ +1$ & $\k\leq R\leq\infty$  \\
          & $L>L^\ast$ & $E>-\k\sqrt{-C}$ & $ +1$ & $\k\leq R\leq\infty$ \\
  \hline
  Deflecting& $L> L^\ast$ & $ E<-\k\sqrt{-C}$ & $ \pm1$ & $R_+\leq R\leq\infty$  \\
  \hline
       & $-L^\ast<L<L^\ast$ & $E>-\k L$ & $ \pm1$ & $\k\leq R\leq R_+$  \\
  Bounded& $ L=L^\ast$ & $-\k L<E<0$ & $ \pm1$ & $\k\leq R\leq R_0$  \\
        & $ L=-L^\ast$ & $E>-\k L$ & $ \pm1$ & $\k\leq R\leq R_0$  \\
        & $ L<-L^\ast$ & $E>-\k L$ & $ \pm1$ & $\k\leq R\leq R_-$  \\
  \hline  \hline
\end{tabular}
\end{table}

\section{General setup}\label{sec:setup}
In this section, we consider the photon emissions from an equatorial source near a Kerr black hole and discuss the near-horizon and near-extremal limits of this problem. In the following, we use $k^\m$ and $p^\mu$ for the four-momentums of an emitter and a photon, respectively, and we use the notation that the quantities with a subscript $``s"$ are for an emitter (source) while the quantities without subscript are for the photons.
In addition, we let $s_{r/\th}=\pm_{r/\th}=\pm1$ denote the orientations of the emitters and let $\s_{r/\th}=\pm_{r/\th}=\pm1$ denote the orientations of the photons.

We consider photon emissions from the equatorial emitters that moves along timelike geodesics. For such emitters we have $\th_s=\pi/2$, thus $Q_s=0$ and
\be
\label{lsstar}
L^\ast_s=l^\ast_s=\frac{2}{\sqrt{3}}M\mu.
\ee
The photon emissions are described by null geodesics, which have
\be
L^\ast=l^\ast=\frac{2}{\sqrt{3}}\sqrt{Q}.
\ee

We will first introduce a pair of local emission angles of the photons in the emitter's local rest frame (LRF), which are expressed in terms of the impact parameters of the photons. Next, we consider the unstable spherical photon orbits, which make up the photon shell and correspond to critical emission angles on the emitter's sky. The critical emission angles will line up into a closed critical curve on the emitter's sky, which allows us to distinguish the photons captured by the central black hole from those escaping to asymptotic infinity \cite{Gates:2020els}. We then define the PEP, whose computation is based on integrating over the interior region of the critical curve, and we find the expression for the MOB.

\subsection{Local emission angles on the emitter's sky}\label{subsec:emittersky}

We use $x^\mu$ with $\mu=0,1,2,3$ to represent both the Boyer-Lindquist coordinates $(t,r,\th,\phi)$ for the Kerr metric \eqref{metric1} and the Bardeen-Horowitz coordinates $(T,R,\th,\Phi)$ for the NHEK \eqref{NKmetric} and near-NHEK \eqref{nNKmetric} metrics.


In terms of the components of the spacetime metric $g_{\m\n}$, the locally nonrotating frame (LNRF) is given by \cite{Bardeen:1972fi}
\be
\label{LNRFdef}
e_{(0)}=\z(\partial_0+\varsigma\partial_3),\qquad
e_{(1)}=\xi_1\partial_1,\qquad
e_{(2)}=\xi_2\partial_2,\qquad
e_{(3)}=\xi_3\partial_3,
\ee
where
\be
\label{LNRFdef1}
\z=\sqrt{\frac{g_{33}}{g_{03}^2-g_{00}g_{33}}},\qquad
\varsigma=-\f{g_{03}}{g_{33}},\qquad
\xi_1=\f{1}{\sqrt{g_{11}}},\qquad
\xi_2=\f{1}{\sqrt{g_{22}}},\qquad
\xi_3=\f{1}{\sqrt{g_{33}}}.
\ee
The LRF of an equatorial emitter can be obtained by performing a Lorentz transformation on the LNRF, which is given by
\begin{subequations}
\label{lrf}
\bea
\label{lrf0}
\s_{[0]}&=&\g[e_{(0)}+v^{(1)}e_{(1)}+v^{(3)}e_{(3)}]|_{x^i=x^i_s},\\
\label{lrf1}
\s_{[1]}&=&\f{1}{v}[v^{(3)}e_{(1)}-v^{(1)}e_{(3)}]|_{x^i=x^i_s}\\
\label{lrf2}
\s_{[2]}&=&e_{(2)}|_{x^i=x^i_s},\\
\label{lrf3}
\s_{[3]}&=&\g[v e_{(0)}+\f{1}{v}(v^{(1)}e_{(1)}+v^{(3)}e_{(3)})]|_{x^i=x^i_s},
\eea
\end{subequations}
where $v$ and $v^{(i)}$ are the 3-velocity and its components of the emitter relative to the LNRF, and $\g$ is the boost factor. For an emitter at $x^\mu_s$ with momentum $k^\mu$, these are given by
\be
\label{velocityandboost}
v^{(i)}=\frac{k^\mu e_\mu^{(i)}}{k^\mu e_\mu^{(0)}}\Bigg|_{x^i=x^i_s},\quad
(i=1,2,3),\qquad
v=\sqrt{(v^{(1)})^2+(v^{(3)})^2},\qquad
\g=\f{1}{\sqrt{1-v^2}}.
\ee

\begin{figure}[h]
  \centering
  \includegraphics[width=7cm]{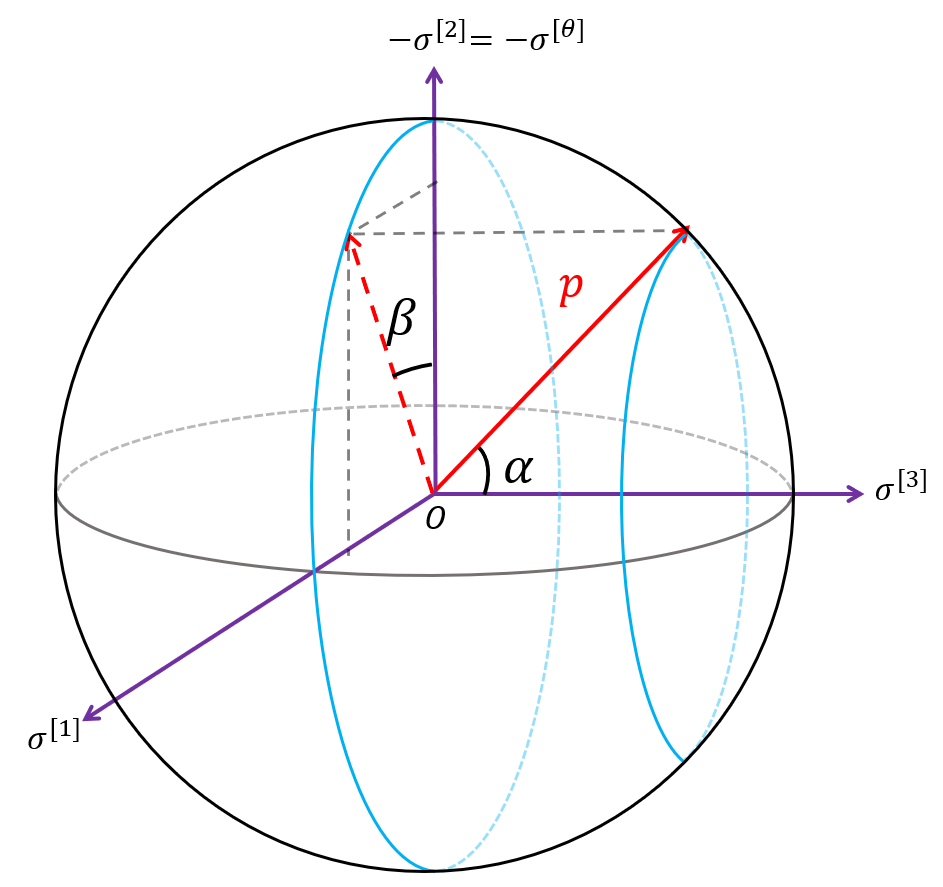}
  \caption{Emitter's sky parameterized by local emission angles $\a$ and $\b$.}
  \label{fig:emittersky}
\end{figure}

In order to study photon emissions from an equatorial emitter, we can define a pair of local emission angles $(\a,\b)$ on the emitter's sky (see Fig.~\ref{fig:emittersky})\cite{Gates:2020els,Ogasawara:2019mir},
\be
\label{localangles}
\a\equiv\arccos\br{\f{p_s^{[3]}}{p_s^{[0]}}}\in[0,\pi],\qquad
\b\equiv\arcsin\br{\f{p_s^{[1]}}{\sqrt{(p_s^{[1]})^2+(p_s^{[2]})^2}}}\in[-\f{\pi}{2},
\f{\pi}{2}],
\ee
where $p^{[a]}_s=p^\mu\s_\mu^{[a]}|_{x^i=x_s^i}$ are the components of the photon's four-momentum in its LRF.
Note that the setup is symmetric about the equatorial plane and only $\s_\theta^2$ appears in the local angles, therefore we may only focus on a half sphere of the emitter's sky by choosing either $\s_\th=1$ or $\s_\th=-1$ in $p_s^{[2]}$. Then the half sphere is uniquely parameterized by $(\a,\b)$.

\subsection{Photon shell and critical emission angles}\label{subsec:criticalangles}
The photon shell consists of unstable spherical photon orbits in the Kerr spacetime, which satisfy
\be
\mathcal{R}(\td r)=\mathcal{R}^\prime(\td r)=0.
\ee
By solving these equations, we can obtain the critical photon impact parameters \cite{Bardeen:1973tla}
\begin{subequations}
\label{kerrcriticalimpacts}
\bea
\label{CriticalL}
\td{\lambda}(\td r)&=&a+\frac{\td r}{a}\br{\td r-\frac{2(\td r^2-2M\td r+a^2) (\td r)}{\td r-M}},\\
\label{CriticalET}
\td{\eta}(\td r)&=&\frac{\td{r}^3}{a^2}\br{\frac{4M(\td r^2-2M\td r+a^2)(\td r)}{(\td r-M)^2}-\td r}.
\eea
\end{subequations}
We have $\td{\eta}(\td r)\geq0$ for the photons crossing the equatorial plane, then these orbits lie in the range $\td r_{1}\leq\td r\leq\td r_{2}$, where
\bea
\label{InnerShell}
\td r_{1}&=&2M\br{1+\cos\pa{\frac{2}{3}\arccos\br{-\frac{a}{M}}}},\\
\label{OuterShell}
\td r_{2}&=&2M\br{1+\cos\pa{\frac{2}{3}\arccos\br{\frac{a}{M}}}}.
\eea
Hereafter, the quantities adorned with a tilde are evaluated for the critical photon emissions.

Plugging \eqref{kerrcriticalimpacts} into \eqref{localangles} gives the critical emission angles which line up into a closed critical curve on an emitter's sky \cite{Gates:2020els}
\be
\label{criticalcurve}
\M{C}=\cu{\pa{\td\a(\td r)\, ,\td\b(\td r)}\bigg|\td r_1<\td r<\td r_2}.
\ee
This critical curve separates the emitter's sky into two distinct regions. To distinguish these regions,
it is useful to introduce the ``direction to the black hole center" on the emitter sky [Eq.~\eqref{localangles}] which corresponds to \cite{Gates:2020els}
\be
\label{dtobh}
\l=\eta=0,\qquad
\s_r=-1.
\ee
Then we call the region containing the ``direction to the black hole center" the captured region, and call the other complementary one the escaping region.

In the extremal limits \eqref{extremalCond}, the intrinsic (near-)NHEK calculations for unstable spherical photon orbits by solving $\M{ R}_{n}(\td R)=\M{R}^\prime_{n}(\td R)=0$ are not sufficient to resolve the whole photon shell described by \eqref{kerrcriticalimpacts}. The reason is that the whole photon shell stretches over various near-horizon limits for a near-extremal black hole \cite{Gates:2020els}.
In the extremal limit $\e \to 0$, the radii of the innermost photon shell \eqref{InnerShell} and the outermost photon shell \eqref{OuterShell} become
\bea
\label{ExtremeInner}
\td r_{1}&=&M\br{1+\e\td R_{1}+\mathcal{O}(\e^2)},\\
\label{ExtremeOuter}
\td r_{2}&=&4M\br{1+\mathcal{O}(\e^2)},
\eea
where $\td R_{1}=2\k/\sqrt{3}$.
We can see that the radius $\td r_1$ is in the near-NHEK $p=1$ region while the radius $\td r_2$ is in the Extreme Kerr $``p=0"$ region [see Eq.~\eqref{BHcoordinates}] \cite{Kapec:2019hro,Gates:2020els}. Therefore, one must rely on the calculations in the Kerr geometry to obtain the whole photon shell.
By expanding the critical impact parameters of the Kerr photon shell [Eqs.~\eqref{kerrcriticalimpacts}] in $\e$ under the transformations \cite{Gates:2020els}
\be
\label{Expansions}
a=M\sqrt{1-(\e\k)^2},\qquad
r_s=M(1+\e^q R_s),\qquad
\td r=M(1+\e^p \td R),
\ee
we can obtain the shifted critical impact parameters $(\td\l_0,\td\eta_0)$.
For the NHEK case we choose $q=2/3$ and for the near-NHEK case we have $q=1$.
We will discuss the expansions for the (near-)NHEK cases in Sec.~\ref{sec:NKemitters}.

\subsection{Photon escaping probabilities and net blueshift}\label{subsec:escapeprobability}
We assume that the emitter emits monochromatic photons isotropically in its LRF \eqref{lrf}.
The redshift factor $g$ and net blueshift $z$ of a photon that reaches to asymptotic infinity are defined by
\be
\label{defredshift}
g\equiv \f{\o}{p_s^{[0]}},\qquad
z\equiv \f{g-1}{g}.
\ee

Let $\mathcal{A}_{e}/\mathcal{A}_{c}$ be the area of the photon escaping/captured region in the emitter's sky of unit radius, respectively.
The photon escaping probability is defined by \cite{Ogasawara:2019mir,Gates:2020els}
\be
\label{defEP}
\mathcal{P}_e\equiv\frac{\mathcal{A}_e}{4\pi}=\frac{1-\mathcal{A}_c}{4\pi}.
\ee
In order to compute the PEP, we
introduce a pair of planar polar coordinates \cite{Gates:2020sdh}
\be
\label{plannercoord}
\rho=\sqrt{2(-\cos\a+1)},\qquad
\varphi=\frac{\pi}{2}+\b,
\ee
such that
the area element $d\mathcal{A}$ on the $(\r,\varphi)$-plane is equal to the area element on the sphere of the emitter's sky $d\Omega$,
\be
\label{AreaEqCond}
d\mathcal{A}=\rho d\rho\wedge d\varphi=\sin\a d\a \wedge d\b=d\Omega.
\ee
Regarding to the position of the direction to the black hole center on the emitter's sky, the area of the interior region of the closed critical curve \eqref{criticalcurve} may correspond to either $\M{A}_e$ or $\M{A}_c$, which can be computed by \cite{Gates:2020els}
\be
\label{AreaCapture}
\mathcal{A}_{in}=\int_{in}\td \rho d\td\rho d\td\varphi
=\int_{in}\frac{1}{2}\td\rho^2d\td\varphi.
\ee


\section{Photon emissions from (near-)NHEK emitters}\label{sec:NKemitters}
In Sec.~\ref{subsec:NKgeodesics}, we have reviewed the (near-)NHEK geometries and the geodesics for the massive particles and for the photons. In this section, we compute the PEP and the MOB for the photon emissions from the emitters at (near-)NHEK radius $R_s$, whose  motions are characterized by $(L_s,E_s,s_r)$. The relevant formulae are just defined in Sec.~\ref{sec:setup}.
We have $q=2/3$ in the Bardeen-Horowitz coordinates [Eq.~\eqref{BHcoordinates}] for the NHEK geometry,
and we have $q=1$ in the Bardeen-Horowitz coordinates for the near-NHEK geometry.

\subsection{Critical emission angles and critical curve }\label{subsec:NKemittersky}
In the last section, we obtained the LRF [Eq.~\eqref{lrf}] and the local emission angles [Eq.~\eqref{localangles}] of an equatorial
emitter at arbitrary radius around a black hole of arbitrary spin.
Now we compute the LRF and the local emission angles for a (near-)NHEK emitter.
At the emitter's position, the LNRF \eqref{LNRFdef} in the (near-)NHEK geometry is given by
\be
\label{nnkLNRF}
e_{(T)}=\f{\p_T-R_s\p_\Phi}{M\sqrt{R_s^2-\k^2}},\qquad
e_{(R)}=\f{\sqrt{R_s^2-\k^2}}{M}\p_R,\qquad
e_{(\th)}=\f{1}{M}\p_\th,\qquad
e_{(\Phi)}=\f{1}{2}\p_\Phi.
\ee
The components of 3-velocity [Eq.~\eqref{velocityandboost}] of an emitter relative to the LNRF are
\be
\label{nnkvelocity}
v^{(R)}=\f{s_r\sqrt{\M{R}_n(R_s)}}{E_s+L_sR_s},\qquad
v^{(\Phi)}=\f{L_s\sqrt{R_s^2-\k^2}}{2(E_s+L_sR_s)},
\ee
where $\M{R}_n$ is defined in \eqref{nnkRpotential}.
Then the local emission angles [Eq.~\eqref{localangles}] of an emitter are given by
\begin{subequations}
\label{nnkangles}
\bea
\label{nnkangleA}
\a(\l_0,\eta_0)&=&\arccos\br{
\frac{-2 v^2 \left(\lambda _0+R_s\right)+2 \s_r \sqrt{\M R_{n0}(R_s)} v^{(R)}+v^{(\Phi) } \sqrt{R_s^2-\kappa ^2}}
{v \left[2 \left(\lambda _0+R_s\right)-2 \s_r v^{(R)}\sqrt{\M R_{n0}(R_s)} -v^{(\Phi) } \sqrt{R_s^2-\kappa ^2}\right]}
},\\
\label{nnkangleB}
\b(\l_0,\eta_0)&=&\arcsin\br{
\frac{2 \s_r v^{(\Phi) }\sqrt{\M R_{n0}(R_s)} -v^{(R)}\sqrt{R_s^2-\kappa ^2}}{\sqrt{4 v^2 \left(R_s^2-\kappa ^2\right) \Theta _{n0}(\f{\pi}{2}) +\left[v^{(R)} \sqrt{R_s^2-\kappa ^2}-2 \s_r v^{(\Phi) }\sqrt{\M R_{n0}(R_s)} \right]{}^2}}
},\,\,\,
\eea
\end{subequations}
where 
\bea
\label{nnknullpotentials}
\M{R}_{n0}(R)&=&\f{\M{R}_{n}(R)}{L^2}=-\eta_0 (R^2-\k^2)+2\l_0 R+\l_0^2+\k^2,\\
\label{nnkThpotential}
\Th_{n0}(\f{\pi}{2})&=&\f{\Th_{n}(\th)}{L^2}\Big|_{\th=\f{\pi}{2}}=\eta_0+\f{3}{4}.
\eea

As discussed in Sec.~\ref{subsec:criticalangles}, the (near-)NHEK critical parameters $(\td\l_0,\td\eta_0)$ [Eq.~\eqref{nkexpandimpacts}] can be obtained by expanding the Kerr critical impact parameters $(\td\l,\td\eta)$ [Eq.~\eqref{kerrcriticalimpacts}]. To be specific, let us consider the expansions \eqref{Expansions} for the emitter's radius $r_s$ with $q=2/3$ or $1$, and for the photon shell radii $\td r$ with $0< p\leq1$.
Then, to the leading order as $\e\rightarrow0$, we obtain
\begin{subequations}
\label{nnkcriticalpara}
\bea
\label{nnkcriticalPlus}
\td\l_0^+=\f{1}{2}\td R^2,\qquad &
\td\eta_0^+=0,\qquad
&\text{for}\quad p=p^+=\f{q}{2},
\\
\label{nnkcriticalMimus}
\td\l_0^-=-\f{\k^2}{\td R},\qquad &
\td\eta_0^-=-\f{\k^2}{\td R^2},\qquad
&\text{for}\quad p=p^-=1,\\
\label{nnkcriticalmiddle}
\td\l_0^{m+}=\infty,\qquad &
\td\eta_0^{m+}=0,\qquad
&\text{for}\quad  p=p^{m+}\in(0,\f{q}{2}), \\
\td\l_0^{m-}=0,\qquad &
\td\eta_0^{m-}=0,\qquad
&\text{for}\quad  p=p^{m-}\in(\f{q}{2},1),
\eea
\end{subequations}
where $\td R(+)\in(0,\infty)$ and $\td R(-)\in[\td R_{1},\infty)$.
Here, the superscripts ``$\pm$" and ``$m$" represent the outer/inner and middle limits of the photon shell, respectively.
The critical emission angles and the critical curve \eqref{criticalcurve} can be obtained by plugging the critical impact parameters \eqref{nnkcriticalpara} into the local emissions angles \eqref{nnkangles} with $\s_r=\text{sign}(\td r-r_s)$.
Note that the limits of $p=p^\pm$ could complete the critical curve, while those of $p=p^{m\pm}$ only contributes the points that connect the $p^\pm$ curves \cite{Gates:2020els,Yan:2021yuo}.

Note also that, if we take the transformation $\td R^2 \rightarrow R_s \td R^2$ for $(\td \a^+,\td \b^+)$, then a NHEK critical curve is independent of $R_s$ once the 3-velocity components \eqref{nnkvelocity} are fixed, and so does the PEP. This is due to the dilational symmetry of the NHEK geometry.
In contrast, a nonzero near-horizon temperature $\k$ in the near-NHEK geometry breaks the dilational symmetry.
Therefore, in contrast with the NHEK case, a near-NHEK critical curve does depend on $R_s$ even when the 3-velocity components \eqref{nnkvelocity} are fixed, and so does the PEP.

\subsection{Photon escaping probability}\label{subsec:NKescape}
Let us now compute the PEP \eqref{defEP} for the photon emissions from a free (near-)NHEK emitter labeled by $(L_s, E_s,s_r)$ [Eq.~\eqref{conservedNK}] at radius $R_s$. To do so, we need to compute the area inside the critical curve \eqref{criticalcurve} in the $(\r,\varphi)$-plane \eqref{plannercoord}.
From \eqref{nkexpandimpacts} and \eqref{dtobh}, the ``direction to the black hole center" is expressed in terms of the shifted critical impact parameters as
\be
\l_0=\infty,\qquad
\eta_0=-\f{3}{4}.
\ee
These parameters correspond to a point in the plane which is contained in the photon captured region.
Then the interior area $\M{A}_{in}$ [Eq.~\eqref{AreaCapture}] can be computed numerically
\be
\label{nkAin}
\M{A}_{in}=\int_{in}\f{1}{2}\td\r^2 d\td\varphi
=\ab{\int_{\td R_1}^\infty \td\r^2(\td\a^-)\f{d\td\varphi(\td\b^-)}{d\td R}d\td R+\int_0^\infty \td\r^2(\td\a^+)\f{d\td\varphi(\td\b^+)}{d\td R}d\td R}.
\ee
Note that we need to integrate over both of the half spheres with  $\s_\th=\pm1$ on the emitter's sky, thus the factor $1/2$ in front of $\td\r^2$ has been canceled out.
We will present the results in Secs.~\ref{sec:dependenceonR} and \ref{motioneffects}, and discuss them in detail there.

\subsection{Maximum observable blueshift}\label{subsec:NKMOB}

As $\e\rightarrow0$, the redshift factor \eqref{defredshift} for the photon emissions from a (near-)NHEK emitter becomes
\be
\label{goflameta}
g(\l_0,\eta_0)
=\f{\sqrt{R_s^2-\k^2}}{\g\br{2(R_s+\l_0)-v^{(\Phi)}\sqrt{R_s^2-\k^2}-2\s_r v^{(R)}\sqrt{\M{R}_{\k0}(R_s)}}}.
\ee
The MOB $z_{\text{mob}}$ is obtained when the redshift factor reaching its maximum value $g_{\text{max}}$.
In order to find the maximum value for the redshift, we analyze the ranges of the impact parameters $(\l_0,\eta_0)$ for the photons that can reach to infinity\footnote{The MOB of the marginally plunging emitters from the ISCO was discussed in \cite{Igata:2021njn} by analyzing the corresponding parameters of the photon emissions.}. We have reviewed the null geodesics in the (near-)NHEK region in Sec.~\ref{subsec:NKgeodesics} and summarized the classifications in Table \ref{table:NKorbits} and Table \ref{table:nNKorbits} for the NHEK and near-NHEK cases, respectively.
For the future-directed photon emissions we have $L^\ast=\f{2}{\sqrt{3}}\sqrt{Q}\geq0$ and $\l_0>-R$, and for the photons arrived at infinity we have\footnote{Note that $\s_r$ will flip from $-1$ to $+1$ at the turning point.} $\s_r=1$.

In the NHEK region we always have $r_s>\td r_1$.
In the near-NHEK region we may either have $r_s>\td r_1$ or have $r_s\leq\td r_1$.
For $r_s>\td r_1$, both the reflecting and the anti-plunging photon trajectories can reach to infinity, while for $r_s\leq\td r_1$, only the anti-plunging photon trajectories can reach to infinity.
Then from Tables \ref{table:NKorbits} and \ref{table:nNKorbits} we have $-1\leq \eta_0\leq0$ and
$\l_0>\l_{0-}(\eta_0) $ for the anti-plunging orbits, and $\l_0<\l_{0-}(\eta_0)$ for the deflecting orbits,
where
\be
\label{lambda0mimus}
\l_{0-}(\eta_0)=-\k\sqrt{-\eta_0}.
\ee
Moreover, the positivity of the potentials \eqref{nnknullpotentials} and \eqref{nnkThpotential}requires
\be
\label{tpointnnk}
\l_0\geq\l_{0t}(\eta_0)= -R_s+ \sqrt{(1+\eta_0)(R_s^2-\k^2)},\qquad
\eta_0\geq-\f{3}{4}.
\ee

For the photon emissions from the infalling emitters with $s_r=-1$ in Eq.~\eqref{nnkvelocity}, we have
$\p_{\eta_0}g >0$ and $\p_{\l_0} g <0$. Note also that we have $\l_0>\l_{0t}$ for $r_s>\td r_1$ and $\l_0>\l_{0-}$ for $r_s\leq\td r_1$. Then we obtain
\begin{align}
\label{nnkgin}
	g_{\text{max},i}=
	\begin{cases}
g_i[\l_{0t}(-3/4),-3/4],	\quad
&\text{if } r_s>\td r_1,	 \\
g_i[\l_{0-}(-3/4),-3/4], \quad
&\text{if } r_s\leq \td r_1,
	\end{cases}
\end{align}
where
\bea
\label{nnkgina}
g_i[\l_{0t}(-3/4),-3/4]&=&\f{1}{\g[1-v^{(\Phi)}]}=\f{\sqrt{3}L_s^\ast \sqrt{R^2_s-\k^2}}{ (2 R_s-\sqrt{R_s^2-\k^2})L_s+2 E_s},\\
\label{gmaxin2}
g_i[\l_{0-}(-3/4),-3/4]&=&\f{\sqrt{R_s^2-\k^2}}{\g\br{2R_s-\sqrt{3}\k-v^{(\Phi)}
\sqrt{R_s^2-\k^2}
-v^{(R)}\sqrt{3R_s^2-4\sqrt{3}R_s\k+4\k^2}}}.\,\,\,\,\,\,\,\,\,\,
\eea
Hereafter, we use subscript $``i/o"$ to represent $s_r=\mp1$ in Eq.~\eqref{nnkvelocity}, respectively.

For the outgoing emitter with $s_r=-1$ in Eq.~\eqref{nnkvelocity}, we also have $\p_{\eta_0}g<0$, and we have $\p_{\l_0}g\gtrless0$ for $\l_0\lessgtr\l_{0c}$,
where
\be
\l_{0c}(\eta_0)=-R_s+\g_R\sqrt{1+\eta_0}\sqrt{R_s^2-\k^2}
\ee
with $\g_R=1/\sqrt{1-(v^{(R)})^2}$. Then we obtain
\begin{align}
\label{nnkgout}
	g_{\text{max},o}=
	\begin{cases}
g_o[\l_{0c}(-3/4),-3/4],	\quad
&\text{if } \l_{0-}<\l_{0c}, 	 \\
g_o[\l_{0-}(-3/4),-3/4], \quad
&\text{if }\l_{0-}>\l_{0c},\,\, (r_s\leq \td r_1\text{ only}),
	\end{cases}
\end{align}
where $g_o[\l_{0-}(-3/4),-3/4]$ has the same form as $g_i[\l_{0-}(-3/4),-3/4]$ [Eq.~\eqref{gmaxin2}] but with $s_r=1$ in $v^{(R)}$, and
\be
\label{nnkgouta}
g_o[\l_{0c}(-3/4),-3/4]=\f{1}{\g\br{1/\g_R-v^{(\Phi)}}}
=\f{\sqrt{3}L_s^\ast}{-L_s+\sqrt{3(L_s^\ast)^2 +L_s^2}}.
\ee

From \eqref{nnkgina} we can see that, for an infalling emitter, an inward radial motion and a retrograde angular motion will redshift the energy of an escaping photon while a prograde angular motion will blueshift the energy of an escaping photon. From \eqref{nnkgouta} we can see that, for an outgoing emitter, the energy of an escaping photon is affected by its velocity in a complex way but is determined only by its angular momentum with a simple relation. We also find that $g_{\text{max},o}$ is always greater than $g_{\text{max},i}$ with the same parameters $(L_s,E_s)$, which means that an outward radial motion will enhance the energy of escaping photons.

In terms of $z_{\text{mob}}$, the MOB for a (near-)NHEK emitter with $r_s>\td r_1$ can be summarized as
\bea
\label{zmobout}
z_{\text{mob},o}&=&1-\f{1}{\sqrt{3}L_s^\ast}\pa{\sqrt{L_s^2
+3(L_s^\ast)^2}-L_s} ,\\
\label{zmobin}
z_{\text{mob},i}&=&1-\f{1}{\sqrt{3}L_s^\ast}\pa{\f{2(L_s R_s+E_s)}{\sqrt{R_s^2-\k^2}}-L_s},
\eea
while the MOB for a (near-)NHEK emitter with $r_s\leq\td r_1$ can be summarized as
\begin{align}
\label{zmobinside}
	z_{\text{mob}}=
	\begin{cases}
1-\f{1}{g[\l_{0-}(-3/4),-3/4]},	\quad
&\text{if } s_r=-1, \text{ or }s_r=1 \,\,\&\,\, \l_{0-}>\l_{0c},	 \\
1-\f{1}{g[\l_{0c}(-3/4),-3/4]}, \quad
&\text{if }s_r=1 \,\,\&\,\, \l_{0-}<\l_{0c}.
	\end{cases}
\end{align}
This is one of our main results and we will analyze the expressions of the MOB for specific emitters in Sec.~\ref{sec:dependenceonR} and plot them in Figs.~\ref{fig:escapeandmob}, \ref{fig:asy1}, \ref{fig:asy2} and \ref{fig:asy3}.

\section{MOB and PEP for different emitters}\label{sec:dependenceonR}
\begin{figure}[hp]
  \centering
  \includegraphics[width=16cm]{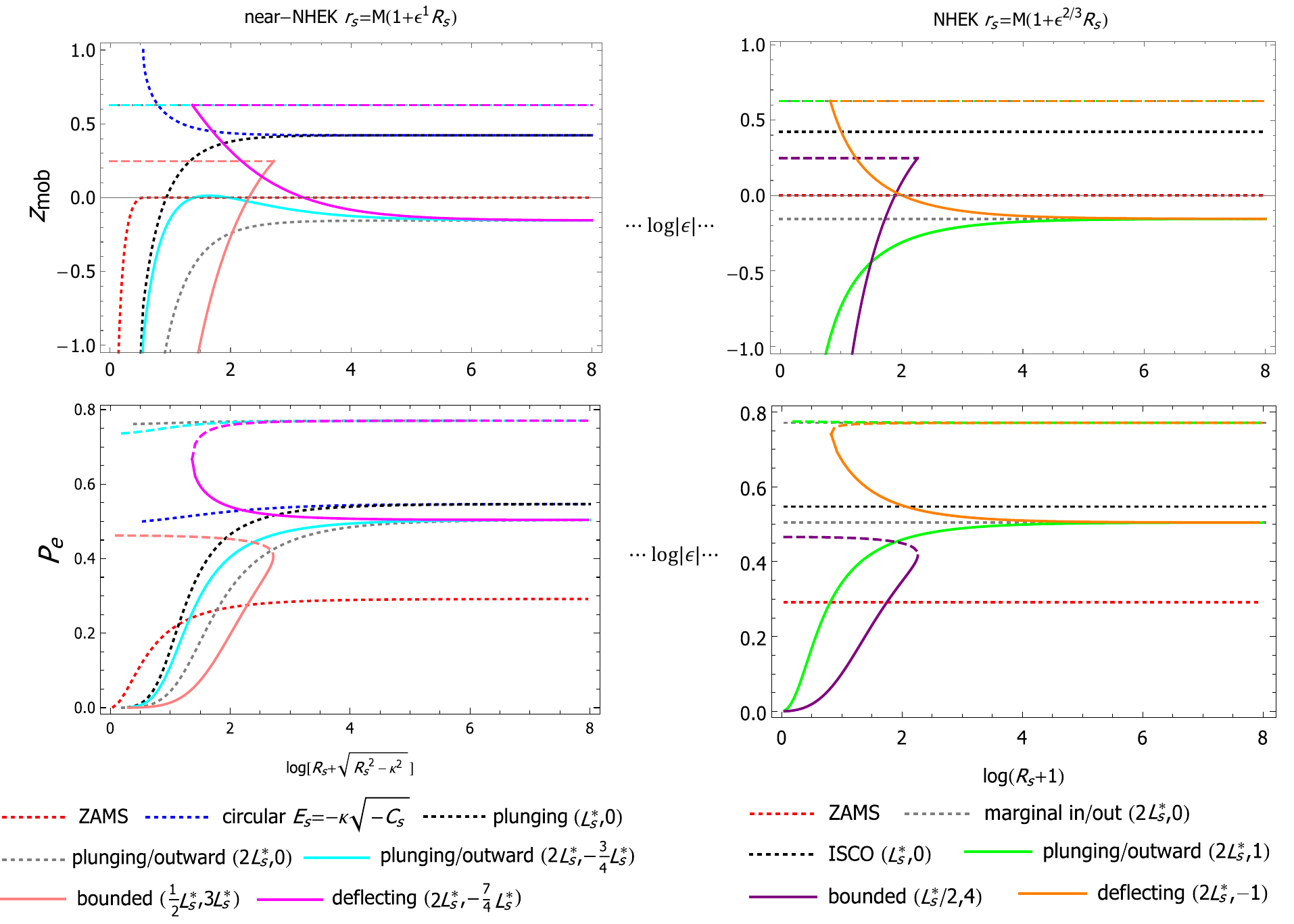}
  \caption{The MOB $z_{\text{mob}}(R_s)$ and the PEP $\M{P}_e(R_s) $ for photon emissions from several representative emitters moving along (near-)NHEK equatorial geodesics. The conserved quantities $(L_s,E_s)$ label the emitters. We show these with the rescaled radial coordinates, representing the proper radial distance \cite{Kapec:2019hro}.
  The dotted curves are used as the reference emitters: the red and blue dotted curves are for ZAMS (which does not follow a geodesic) and the emitters at circular geodesics, respectively, the black and gray dotted curves are for the emitters on the marginal (plunging/anti-plunging/deflecting) orbits with $s_r=\pm1$.
  The dashed and solid curves are for the emitters on the outgoing ($s_r=+1$) and ingoing ($s_r=-1$) orbits, respectively.  We have set $M=1$, $\mu=1$ and $\k=1$ (for the near-NHEK case).}
  \label{fig:escapeandmob}
\end{figure}

\begin{figure}[hp]
  \centering
  \includegraphics[width=16.5cm]{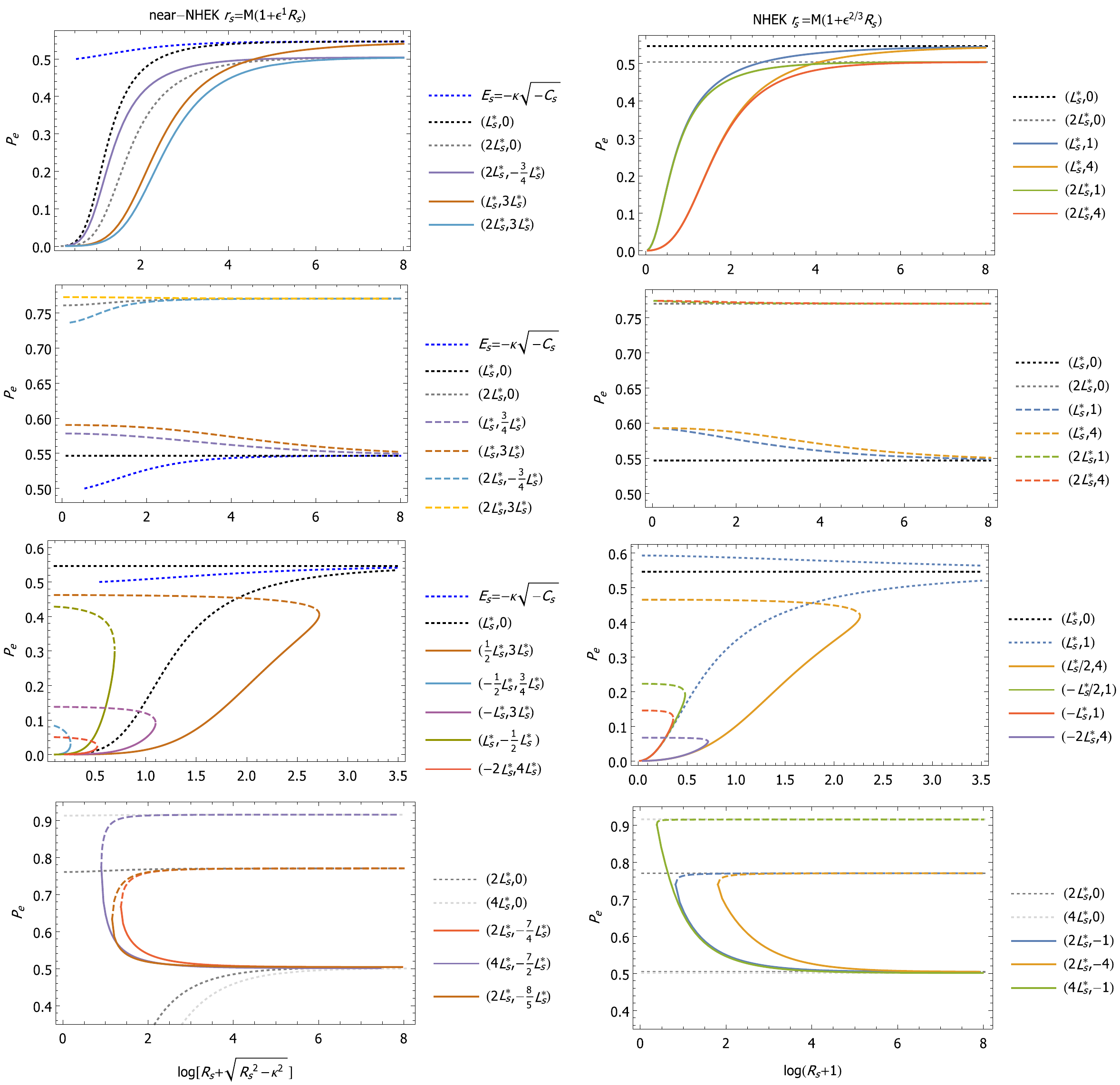}
  \caption{Top to bottom: examples for the PEPs for the emitters along the plunging, the anti-plunging, the bounded and the deflecting orbits, respectively. The conserved quantities $(L_s,E_s)$ [Eq.~\eqref{conservedNK}] label the emitters. The dashed and solid curves are for the sources on the outgoing ($s_r=+1$) and infalling  ($s_r=-1$) orbits, respectively. The dotted curves are displayed as the references.  }
  \label{fig:escapetogether}
\end{figure}


In Fig.~\ref{fig:escapeandmob}, we display the results of $z_{\text{mob}}(R_s)$ and $\M{P}_e(R_s)$ of the escaping photons emitted from different orbits labeled by $(L_s,E_s,s_r)$.
We can see that the behaviors of PEP are closely related to that of MOB.
In addition, we also present more examples of the PEPs in Fig.~\ref{fig:escapetogether} for the emitters along the plunging, the anti-plunging, the bounded and the deflecting orbits, respectively.
We  take an ZAMS as the reference emitter which has $v^{(R)}=0$ and $v^{(\Phi)}=0$.
Note that the ZAMS does not follow a geodesic and we will use the term ``orbits" only for the geodesics. We take the emitters moving on circular orbits as the references as well.
We use red and blue dotted curves to show the results for ZAMSs and circular  orbits, respectively. The dotted curves with other colors are used as additional reference emitters whose motions are described in the  plots.
The dashed and solid curves are, respectively, for the infalling ($s_r=-1$) and the outgoing ($s_r=1$) emitters along several orbits labeled by $(L_s,E_s)$.
We show these curves with the rescaled radial coordinates
\be
s_{\text{NK}}=\log(R+1),\qquad
s_{\text{nNK}}=\log\br{R+\sqrt{R^2-\k^2}},
\ee
in terms of which, the proper radial distances in the NHEK and near-NHEK regions are \cite{Kapec:2019hro}
\be
ds_{\text{NK}}(\bar{R}_1,\bar{R}_2)=s_{\text{NK2}}-s_{\text{NK1}},\qquad
ds_{\text{nNK}}(R_1,R_2)=s_{\text{nNK2}}-s_{\text{nNK1}},\qquad
\bar{R}=R+1,
\ee
while the proper radial distance between the NHEK and near-NHEK regions scales as $\log|\e|$ \cite{Bardeen:1972fi,Kapec:2019hro}.
Note that from the perspective of near-NHEK geometry, the $R_{\text{nNK}}\rightarrow\infty$ limit glues onto the NHEK region. 
Here the subscripts ``NK" and ``nNK" represent the quantities in the NHEK and near-NHEK geometries, respectively.
In practice, we choose $M=1$, $\mu=1$ for the emitter's mass, and $\kappa=1$ for the near-NHEK case.

Before describing the main results of this paper, we recall the relevant results for photon emissions from ZAMS and from the emitters on circular geodesics.
For photon emissions from a ZAMS, we have
\begin{align}
	z_{\text{mob}}(R_s)|_{\text{ZAMS}}=
	\begin{cases}
0,	\quad
&\text{if } r_s>\td r_1,	 \\
1-\f{2R_s-\sqrt{3}\k}{\sqrt{R_s^2-\k^2}}, \quad
&\text{if } r_s\leq \td r_1.
	\end{cases}
\end{align}
It was found in \cite{Ogasawara:2019mir,Yan:2021yuo} that the PEP for a NHEK ZAMS is about $29\%$ and the PEP for a near-NHEK ZAMS decreases as the source radius decreases from near-NHEK infinity towards horizon radius and reaches approximately $13\%$ at $\td r_1$.
For photon emissions from the emitters on circular orbits in the range $r_s>\td r_1$ ($\td r_1$ is also the last allowed radius for a circular orbit \cite{Bardeen:1972fi}, at which $L_s\rightarrow\infty$),
we have
\begin{align}
	z_{\text{mob}}(R_s)|_{\text{cir}}=
	\begin{cases}
1-\f{1}{\sqrt{3}},	\quad
&\text{if } r_s=M(1+\e^{\f{2}{3}} R_s),\quad (\text{ISCO}), \\
1-\f{1}{\sqrt{3}L_s^\ast}\pa{\f{2(L_s R_s-\k\sqrt{-C_s})}{\sqrt{R_s^2-\k^2}}-L_s}, \quad
&\text{if } r_s=M(1+\e R_s),\quad R_s>\td R_1.\,\,\,\,\,\,\,\,
	\end{cases}
\end{align}
All circular orbits in the NHEK region are the ISCO while the ones in the near-NHEK region are unstable.
For a given unstable circular orbit in the near-NHEK region,
the orbital radius is uniquely determined by $R_c=\k L_s/\sqrt{-C_s}$, then we obtain
$z_{\text{cir},u}=z_{\text{mob}}(R_c)|_{\text{cir}}=z_{\text{mob},o}$.
Here, the subscript $``u"$ represents ``unstable".
For photon emissions from the ISCO\footnote{For photon emissions from stable circular orbits outside the ISCO of an extremal Kerr black hole, the PEP decreases monotonously as the orbital radius decreases and reaches approximately $55\%$ at the ISCO \cite{Igata:2019hkz,Gates:2020els}.}, the PEP is about $55\%$  \cite{Igata:2019hkz,Gates:2020els}.
We find that, for photon emission from unstable circular orbits, the PEP decreases from $\sim55\%$ to $\sim50\%$ as the orbital radius decreases from the one of ISCO towards $\td r_1$.

Let us first look at the MOB (see Fig.~\ref{fig:escapeandmob}) for a (near-)NHEK emitter labeled by $(L_s,E_s,s_r)$.
For photon emissions from an outgoing emitter $(L_s,E_s,+1)$, the MOB is a constant along a given orbit, i.e., the MOB does not depend on $R_s$.
For photon emissions from a marginal infalling emitter which exists only in the NHEK region, the MOB does not depend on $R_s$ as well and we have
\be
\label{zmarginal}
z_{\text{mar},i}=z_{\text{mob},i}|_{\text{marginal}}=1-\f{L_s}{\sqrt{3}L_s^\ast}.
\ee
For photon emissions from a bound/deflecting emitter along the infalling part of its orbit,
$z_{\text{mob},i}(R_s)$ decreases/increases monotonously as the emitter moves from (near-)NHEK infinity towards horizon.
For photon emissions from the plunging emitters, the behavior of $z_{\text{mob},i}(R_s)$ between the NHEK and near-NHEK cases are different:
for photon emissions from a NHEK plunging emitter or from a near-NHEK plunging emitter with $E_s\geq0$, $z_{\text{mob},i}(R_s)$ decreases monotonously as the emitter moves from (near-)NHEK infinity towards horizon; while for photon emissions from a near-NHEK plunging emitter with $-\k\sqrt{-C_s}<E_s<0$, $z_{\text{mob},i}(R_s)$ increases at the beginning as the emitter leaves from the near-NHEK infinity until reaching an extreme value at
\be
\label{Rextreme}
R_e=-\f{\k^2L_s}{E_s},
\ee
then $z_{\text{mob},i}(R_s)$ begins to decrease as the emitter falls towards the horizon, and at $R_e$ we find
$ z_{\text{mob},i}= z_{\text{mob},o}=z_{\text{cir},u}$ as $ E_s\rightarrow-\k\sqrt{-C_s}$.
Moreover,
at the (near-)NHEK infinity, we have $z_{\text{mob},i}(R_\infty)=z_{\text{mar},i}$ for a plunging or a deflecting emitter,
and at the the turning point of a bounded or a deflecting orbit, we have $z_{\text{mob},i}=z_{\text{mob},o}$.
These are summarized in Tables \ref{table:NKbehavior2} and \ref{table:nNKbehavior2}.


Next, we turn to the PEP (see Figs.~\ref{fig:escapeandmob} and \ref{fig:escapetogether}) for a (near-)NHEK emitter labeled by $(L_s,E_s,s_r)$.
For photon emissions from a plunging emitter, $\M{P}_e(R_s)$ decreases monotonously along its orbit as the emitter moves from (near-)NHEK infinity towards the horizon.
For photon emissions from a bounded emitter, $\M{P}_e(R_s)$ decreases monotonously along its orbit as the emitter moves outward from the horizon towards the turning point and then bounces off back to the horizon.
For photon emissions from a deflecting emitter, $\M{P}_e(R_s)$ increases monotonously along its orbit as the emitter moves inward from (near-)NHEK infinity towards the turning point and then bounces off back towards (near-)NHEK infinity.
For photon emissions from the anti-plunging emitters, there are differences between the NHEK and near-NHEK cases:
for a NHEK anti-plunging emitter and for a near-NHEK anti-plunging emitter with $E_s>\k\sqrt{-C_s}$, $\M{P}_e(R_s)$ decreases monotonously along its orbit from the horizon towards infinity; while for a near-NHEK anti-plunging emitter with $E_s<\ab{\k\sqrt{-C_s}}$, $\M{P}_e(R_s)$ increases monotonously along its orbit from the horizon towards infinity.
Furthermore,
we find that for all (anti-plunging and deflecting) emitters that can eventually leave from the NHEK infinity, we always have $\M{P}_e>50\%$ while for all plunging emitters that fall into the black hole we always have $\M{P}_e<55\%$. In addition, for the bounded emitters in the (near-)NHEK region, we always have $\M{P}_e<59\%$.
These are also summarized in Tables \ref{table:NKbehavior2} and \ref{table:nNKbehavior2}.

\section{Influence of the emitter's proper motion}\label{motioneffects}
From Fig.~\ref{fig:escapeandmob}, we can also see that the PEP and MOB are clearly affected by the proper motion of an emitter. For an emitter with proper motion labeled by $(L_s,E_s,s_r)$, the behaviors of PEP and MOB along its orbit are clarified in the previous section. Next, we consider the behaviors of PEP and MOB as the orbital constants $(L_s,E_s,s_r)$ vary. To do so, we study the functions $z_{\text{mob}}(L_s,E_s,s_r)$ and $\M{P}_e(L_s,E_s,s_r)$ at several representative points along each emitter's orbit.

In the last section, we find that
$z_{\text{mob},o}$ [Eq.~\eqref{zmobout}] depends only on $L_s$ and tends to $1$ (infinite blueshift) as $L_s\rightarrow\infty$ while tends to $-\infty$ (infinite redshift) as $L_s\rightarrow-\infty$. 
Moreover, $z_{\text{mob},i}$ [Eq.~\eqref{zmobin}] of an infalling emitter is always less than that of an outgoing one with the same parameters $(L_s,E_s)$, and $z_{\text{mob},i}$ depends on both $L_s$ and $E_s$.
We now consider only  photon emissions with $z_{\text{mob}}\geq0$ and make a cutoff along a given orbit (if relevant) at $R_z$ where $z_{\text{mob}}(R_z)=0$. In other words, we focus on the prograde orbits ($L_s>0$) for the emitters since
no photon emissions from the retrograde emitters ($L_s<0$) are blueshifted.

\subsection{NHEK emitters}
\begin{figure}[h]
  \centering
  \includegraphics[width=16cm]{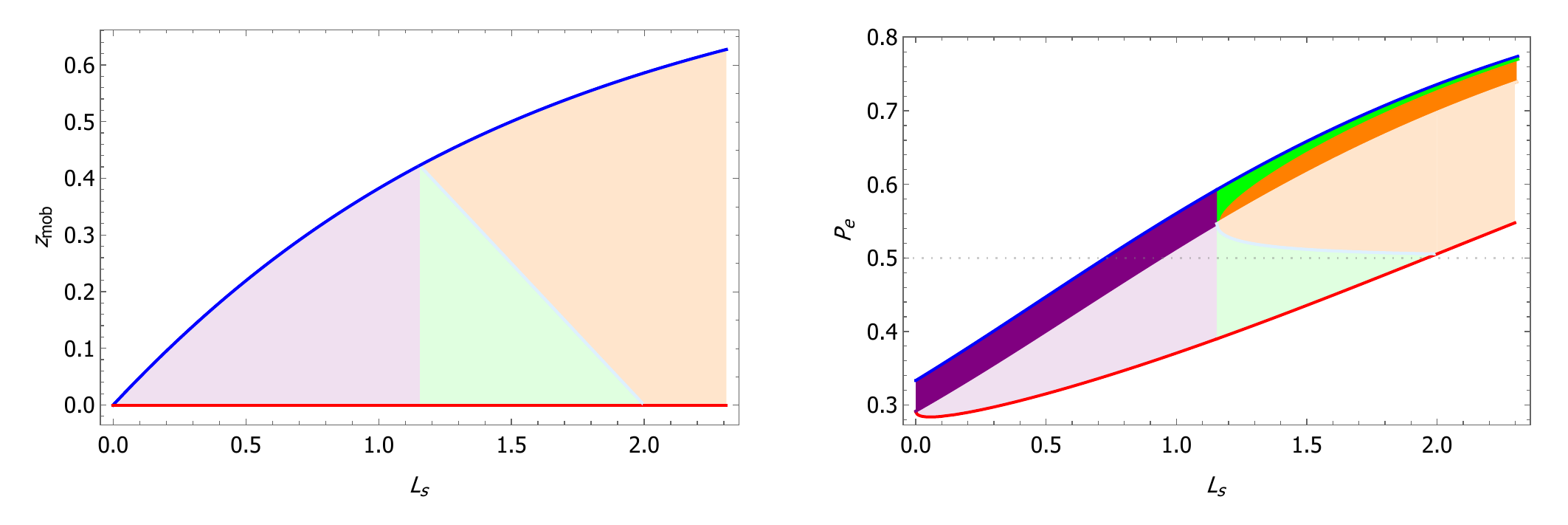}
  \caption{Values of $z_{\text{mob}}(L_s)$ and $\M{P}_e(L_s)$ between the endpoints of the part of each NHEK equatorial orbit with $z_{\text{mob}}\geq0$. The radial ranges are illustrated explicitly in Table \ref{table:NKbehavior2}. The blue solid line in the left is for $z_{\text{mob},o}$ and the red solid line has $z_{\text{mob}}=0$. The lightpurple, lightgreen and lightorange ranges are for photon emissions from the infalling ($s_r=-1$) bounded, plunging and deflecting emitters, respectively, while the purple, green and orange ranges are for photon emissions from the outgoing ($s_r=+1$) bounded, anti-plunging and deflecting emitters, respectively.  }
  \label{fig:asy1}
\end{figure}
Table \ref{table:NKbehavior1} shows the components of the 3-velocities of the NHEK emitters relative to the ZAMS at the endpoints of these emitters' orbits or at the cutoff point
\be
R_z=\f{2 E_s}{\sqrt{3}L_s^\ast-L_s}.
\ee
We find that the velocities are independent of $E_s$ at these points.
And a change of $E_s$ will only extend or shrink the range with $z_{\text{mob}}\geq0$  by shifting the points $R_z$ or $R_+$.
As $\ab{E_s}$ increases, $R_+$ and $R_z$ increase as well. For a bounded orbit,  the radial range $R\in(0,R_+)$ is extended while the entire region $R\in(R_z,R_+)$ appears farther away from the horizon.
For a deflecting orbit, the radial range $R\in(R_+,\infty)$ is shrunk while the entire region $R\in(R_+,R_z)$ appears closer to the horizon.
For a plunging emitter, the radial range $R\in(R_z,\infty)$ is shrunk.
We find that the PEP and MOB are also independent of $E_s$ at the endpoints or at the cutoff point.

\begin{table}[h]
\centering
\caption{The 3-velocity components at the endpoints ($0$, $R_+$ or $R_\infty$) or the cutoff point ($R_z$) of a NHEK equatorial orbit.}
  \label{table:NKbehavior1}
\begin{tabular}{c c c c c }
  \hline  \hline
   $R_s$ & $0$ & $R_z$& $R_+$   & $R_\infty$ \\
  \hline
   $v^{(R)}$ & $\pm1$ & $\pm \sqrt{2\sqrt{3}L_s^\ast L_s}/(\sqrt{3}L_s^\ast+L_s)$  & $0$   &  $\pm \sqrt{-C_s}/L_s$ \\
  \hline
   $v^{(\Phi)}$ & $0$ & $L_s/(\sqrt{3}L_s^\ast+L_s) $  & $L_s/\sqrt{3(L_s^\ast)^2+L_s^2}$  &  $1/2 $\\
  \hline  \hline
\end{tabular}
\end{table}

\begin{table}[h]
\centering
\caption{Behaviors of $z_{\text{mob}}(L_s,E_s,s_r)$ and $\M{P}_e(L_s,E_s,s_r)$ between the endpoints of the part of each NHEK equatorial orbit with $z_{\text{mob}}\geq0$, where $ z_{\text{mar},i}$ and $ z_{\text{mob},o}$ are given in Eqs.~\eqref{zmarginal} and \eqref{zmobout}, respectively.}
  \label{table:NKbehavior2}
\begin{tabular}{c c c c c c}
  \hline  \hline
   $L_s$ & $E_s$ & $s_r$ & $z_{\text{mob}}$ & $\M{P}_e$ & motion \\
  \hline
   $0<L_s<L_s^\ast$ & $E_s>0$&  $+$ & $z_{\text{mob},o}$ & $\M{P}_e(R_+)<\M{P}_e<\M{P}_e(0)$ & bounded \\
   \hline
   $0<L_s<L_s^\ast$ & $E_s>0$&  $-$ & $(0,\,\, z_{\text{mob},o})$ & $\M{P}_e(R_z)<\M{P}_e<\M{P}_e(R_+)$ & bounded  \\
   \hline
      $L_s=L_s^\ast$ & $E_s=0$&  $\shortmid$ & $1-1/\sqrt{3}$ & $\sim 55\%$ &ISCO \\
   \hline
   $L_s\geq L_s^\ast$ & $E_s>0$&  $+$ & $z_{\text{mob},o}$ & $\M{P}_e(R_\infty)<\M{P}_e<\M{P}_e(0)$ &anti-plunging \\
    \hline
    $L_s\geq L_s^\ast$ & $E_s>0$&  $-$ & $(0,\,\, z_{\text{mar},i})$ & $\M{P}_e(R_z)<\M{P}_e<\M{P}_e(R_\infty)$ &plunging\\
    \hline
    $L_s>L_s^\ast$ & $E_s=0$&  $\pm$ & $z_{\text{mob},o}/z_{\text{mar},i}$ & $\M{P}_{\text{mar},o}/\M{P}_{\text{mar},i} $ &marginal \\
    \hline
   $L_s>L_s^\ast$ & $E_s<0$&  $+$ & $z_{\text{mob},o}$ & $\M{P}_e(R_+)<\M{P}_e<\M{P}_e(R_\infty)$ &deflecting \\
    \hline
   $L_s^\ast<L_s<2$ & $E_s<0$&  $-$ & $(0,\,\, z_{\text{mob},o})$ & $\M{P}_e(R_z)<\M{P}_e<\M{P}_e(R_+)$& deflecting \\
    \hline
   $L_s>2$ & $E_s<0$&  $-$ & $ (z_{\text{mar},i},\,\, z_{\text{mob},o})$ & $\M{P}_e(R_\infty)<\M{P}_e<\M{P}_e(R_+)$ &deflecting  \\
  \hline  \hline
\end{tabular}
\end{table}

In Table \ref{table:NKbehavior2}, we display distinct behaviors of $z_{\text{mob}}(L_s,E_s,s_r)$ and $\M{P}_e(L_s,E_s,s_r)$ between the endpoints of the part of an emitter's orbit with $z_{\text{mob}}\geq0$.
Meanwhile, we display in Fig.~\ref{fig:asy1} the corresponding values of $z_{\text{mob}}$ and $\M{P}_e$.
From Fig.~\ref{fig:asy1}, we can see that the maximum PEP is increased monotonously as $L_s$ is increased from $0$.
For $L_s=0$, the maximum PEP is about $33\%$ and the PEP at $R_z$ is about $29\%$.
For $L_s=L_s^\ast$, the maximum PEP for the anti-plunging emitters is about $59\%$ and the minimum/maximum PEP for the anti-plunging/plunging emitters is about $55\%$, and the PEP for the plunging emitters at $R_z$ is about $39\%$.
For $L_s=\sqrt{3}L_s^\ast$, the PEP for the anti-plunging emitters is in a narrow range around $73\%$ and the PEP for the plunging emitters in $(R_z,R_\infty)$ is about $50\%$. However, as $L_s\rightarrow \sqrt{3}L_s^\ast$, the cutoff point $R_z$ tends to infinity and thus the radial region with $z_{\text{mob}}\geq0$ is actually shrunk to zero.
For $L_s>\sqrt{3}L_s^\ast$, photon emissions from the anti-plunging and deflecting orbits always have a PEP greater than $50\%$ while photon emissions from the plunging orbits never have a PEP greater than $50\%$.

As expected, an anti-plunging or deflecting emitter can be observed ( i.e, with considerable PEP and carrying enough energy at infinity) by a distant observer, since the emitter itself can also reach to asymptotic far region. However, for a plunging emitter that eventually drops into the central black hole, it is interesting to ask to what depth in the near-horizon throat can the emitter still be observed.
We find that, among all plunging emitters, the one with critical parameters $( L_s^\ast, 0)$ has a maximum PEP and the range of $(R_z,R_\infty)$ extends to the entire NHEK region.

\subsection{Near-NHEK emitters}
Unlike those for the NHEK emitters, the 3-velocity components relative to the ZAMS for an near-NHEK plunging emitter with $-\k\sqrt{-C_s}<E_s<0$ are no longer monotonous along its orbits, and neither does the MOB of such an emitter. Instead, there is an extreme velocity along each of these orbits at the critical point $R_e$ [Eq.~\eqref{Rextreme}].
Table \ref{table:nNKbehavior1} shows the components of an emitter's 3-velocity at the endpoints, the critical point of the emitter's orbit or the cutoff point
\be
R_z=\f{2 E_s L_s\pm(\sqrt{3}L_s^\ast +L_s)\sqrt{E_s^2+\k^2( C_s+\f{\sqrt{3}}{2}L_s^\ast L_s)}}{2C_s+ \sqrt{3}L_s^\ast L_s},
\ee
where the plus/minus sign is taken as $L_s\lessgtr\sqrt{3}L_s^\ast$, respectively. In the table, the 3-velocity components at the critical points are
\be
v^{(R)}(R_e)=-\sqrt{(E_s^2+\k^2 C_s)/(E_s^2-\k^2 L_s^2)}, \hspace{3ex} v^{(\Phi)}(R_e)=\k L_s/\sqrt{4(E_s^2-\k^2 L_s^2)}.
\ee
Note that, as $E_s\rightarrow-\k\sqrt{-C_s}$, the 3-velocity components obtained at $R_e$ have the same forms as those obtained at the turning point $R_+$ of a bounded or deflecting orbit.
We find that the velocities do not depend on $E_s$ at the endpoints ($\k$, $R_+$ or $R_\infty$) and at the cutoff point $R_z$, but do depend on $E_s$ at the critical point $R_e$.
As in the NHEK cases, a change of $E_s$ will extend or shrink the range with $z_{\text{mob}}\geq0$ in a similar way.
However, for the near-NHEK case, the PEP and MOB depend on $E_s$ at the endpoints, the cutoff point and the critical point. And in the limit $\ab{E_s}\rightarrow\infty$, the PEP and MOB tend to the values of the corresponding NHEK cases.

\begin{table}[h]
\centering
 \caption{The 3-velocity components at the endpoints ($\k$, $R_+$ or $R_\infty$), the cutoff point ($R_z$), or the critical point ($R_e$) of a near-NHEK equatorial orbit.}
  \label{table:nNKbehavior1}
\begin{tabular}{c c c c c c}
  \hline  \hline
   $R_s$ & $\k$ & $R_z$& $R_e$ & $R_+$   & $R_\infty$ \\
  \hline
   $v^{(R)}$ & $\pm1$ & $\pm \sqrt{2\sqrt{3}L_s^\ast L_s}/(\sqrt{3}L_s^\ast+L_s)$   & $v^{(R)}(R_e)$  & $0$   &  $\pm \sqrt{-C_s}/L_s$ \\
  \hline
   $v^{(\Phi)}$ & $0$ & $L_s/(\sqrt{3}L_s^\ast+L_s)$  & $v^{(\Phi)}(R_e)$  & $L_s/\sqrt{3(L_s^\ast)^2+L_s^2}$  &  $1/2 $\\
  \hline  \hline
\end{tabular}
 \end{table}

\begin{table}[h]
\centering
\caption{Behaviors of $z_{\text{mob}}(L_s,E_s,s_r)$ and $\M{P}_e(L_s,E_s,s_r)$ between the endpoints of the part of each near-NHEK equatorial orbit with $z_{\text{mob}}\geq0$.}
  \label{table:nNKbehavior2}
\begin{tabular}{c c c c c c c}
  \hline  \hline
   $L_s$ & $E_s$ & $s_r$ & $z_{\text{mob}}$ & $\M{P}_e$& motion  \\
  \hline
   $L_0<L_s<L_s^\ast$ & $E_s>-\k L_s$&  $+$ & $z_{\text{mob},o}$ & $\M{P}_e(R_+)<\M{P}_e<\M{P}_e(\k)$ & bounded \\
   \hline
   $L_0<L_s<L_s^\ast$ & $E_s>-\k L_s$&  $-$ & $(0,\,\, z_{\text{mob},o})$ & $\M{P}_e(R_z)<\M{P}_e<\M{P}_e(R_+)$ & bounded\\
    \hline
   $L_s=L_s^\ast$ & $E_s\geq0$&  $+$ & $z_{\text{mob},o}$ & $\M{P}_e(R_\infty)<\M{P}_e<\M{P}_e(\k)$ &anti-plunging \\
  $L_s>L_s^\ast$ & $E_s>\k\sqrt{-C_s}$&  &  &  &  \\
    \hline
    $L_s>L_s^\ast$ & $E_s<\ab{\k\sqrt{-C_s}}$&  $+$ & $z_{\text{mob},o}$ & $\M{P}_e(\k)<\M{P}_e<\M{P}_e(R_\infty)$ &anti-plunging \\
    \hline
  $L_s>L_s^\ast$ & $-\k\sqrt{-C_s}<E_s<0$&  $-$ & $(0,\,\,z_{\text{mob},i}(R_e))$ & $\M{P}_e(R_z)<\M{P}_e<\M{P}_e(R_\infty)$ &plunging \\
    \hline
    $L_s\geq L_s^\ast$ & $E_s\geq0$&  $-$ & $(0,\,\,  z_{\text{mar},i})$ & $\M{P}_e(R_z)<\M{P}_e<\M{P}_e(R_\infty)$ &plunging\\
    \hline
 $L_s>L_s^\ast$ & $E_s=-\k\sqrt{-C_s}$&  $\shortmid$ & $z_{\text{mob},o}$ & $\M{P}_{\text{cir},u}$& circular  \\
    \hline
   $L_s>L_s^\ast$ & $E_s<-\k\sqrt{-C_s}$&  $+$ & $z_{\text{mob},o}$ & $\M{P}_e(R_+)<\M{P}_e<\M{P}_e(R_\infty)$& deflecting  \\
    \hline
   $L_s^\ast<L_s<2$ & $E_s<-\k\sqrt{-C_s}$&  $-$ & $(0,\,\, z_{\text{mob},o})$ & $\M{P}_e(R_z)<\M{P}_e<\M{P}_e(R_+)$ &deflecting  \\
    \hline
   $L_s>2$ & $E_s<-\k\sqrt{-C_s}$&  $-$ & $(z_{\text{mar},i},\,\, z_{\text{mob},o})$ & $\M{P}_e(R_\infty)<\M{P}_e<\M{P}_e(R_+)$ & deflecting \\
  \hline  \hline
\end{tabular}
\end{table}

\begin{figure}[hp]
  \centering
  \includegraphics[width=16cm]{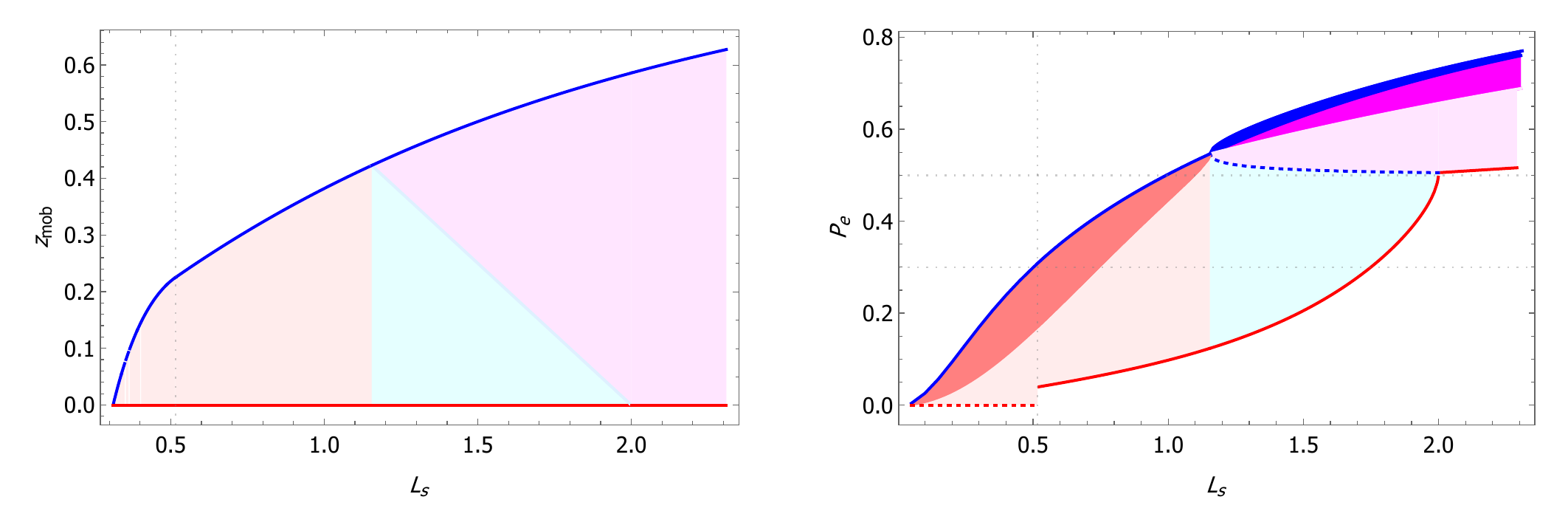}
  \caption{Values of $z_{\text{mob}}(L_s)$ and $\M{P}_e(L_s)$ between the endpoints of the part of each near-NHEK equatorial orbit with $z_{\text{mob}}\geq0$. The ranges are illustrated explicitly in Table \ref{table:nNKbehavior2}.
  For the emitters with $L_0<L_s<\f{L_s^\ast}{\sqrt{5}}$ and $E_s=0$, the MOB $z_\text{mob}$ is obtained for photon emissions from $\td r<\td r_1$, otherwise, it is given in Eqs.~\eqref{zmobout} and \eqref{zmobin}.
  We have chosen $E_s=0$ for the bounded, the plunging and the anti-plunging emitters, and have chosen $E_s=-\k L_s$ for the deflecting emitters.
  The solid blue line in the left is for $z_{\text{mob},o}$ and the solid red line has $z_{\text{mob}}=0$, while the dotted red line in the right has $z_{\text{mob}}\rightarrow-\infty$.
  The lightpink, lightblue and lightmagenta ranges are for photon emissions from the infalling ($s_r=-1$) bounded, plunging and deflecting emitters, respectively, while the pink and magenta ranges are for photon emissions from the outgoing ($s_r=+1$) bounded and deflecting emitters, respectively, and the blue range are the overlap range for photon emissions from the anti-plunging and outgoing deflecting emitters.
  }
  \label{fig:asy2}
\end{figure}

\begin{figure}[hp]
  \centering
  \includegraphics[width=16cm]{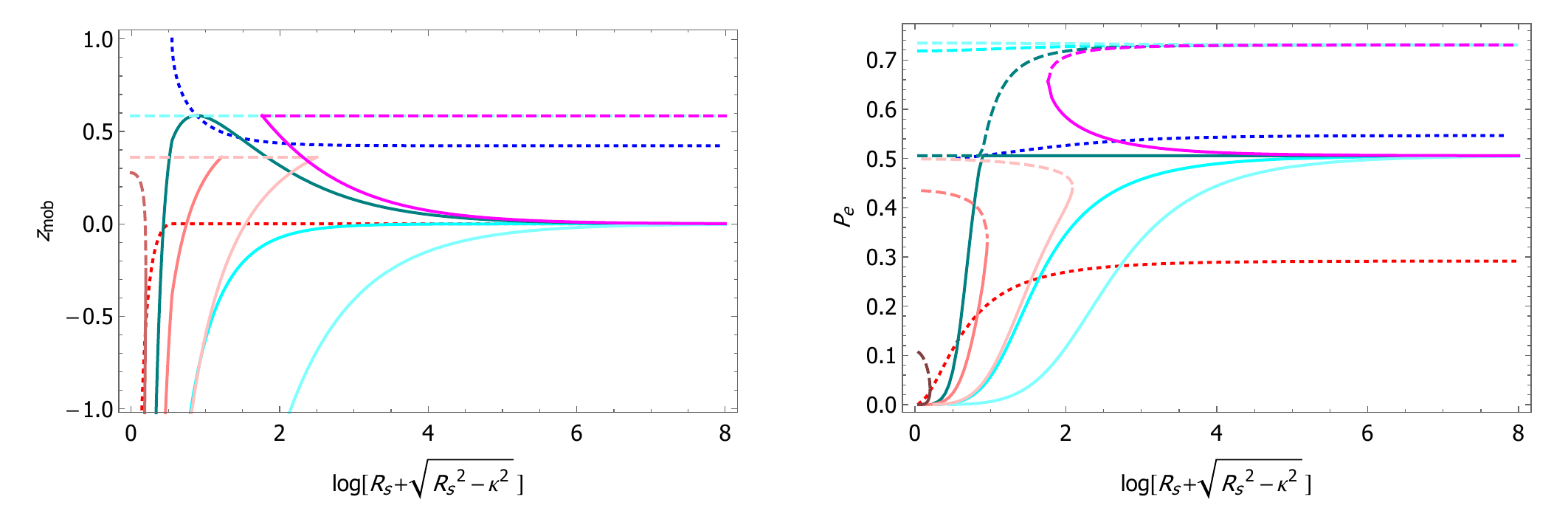}
  \caption{$z_{\text{mob}}(R_s)$ and $\M{P}_e(R_s)$ for photon emissions from several near-NHEK emitters. The red dotted curve is for ZMAS, and the blue dotted curve is for the emitters on circular geodesic orbits. The dashed curves are for $s_r=+1$ and the solid curves are for $s_r=-1$. The darkcyan, cyan and lightcyan have $L_s=\sqrt{3}L_s^\ast$ and $E_s=-\k\sqrt{-C_s}$, $0$ and $2\sqrt{3}L_s^\ast$, respectively. The magenta curve has $L_s=\sqrt{3}L_s^\ast$ and $E_s=-\k L_s$. The darkpink, pink and lightpink curves have $L_s=\f{4}{5}L_s^\ast$ and $E_s=-\f{(8-\sqrt{3})}{10}L_s^\ast$, $0$ and $\f{\sqrt{3}}{2}L_s^\ast$, respectively. }
  \label{fig:asy3}
\end{figure}

Table \ref{table:nNKbehavior2} displays distinct behaviors of $z_{\text{mob}}$ and $\M{P}_e$ between the endpoints of the part of an emitter's orbit with $z_{\text{mob}}\geq0$. $z_{\text{mob},o}$ is given in Eq.~\eqref{zmobout} if $r_s\geq\td r_1$ or in Eq.~\eqref{zmobinside} if $r_s<\td r_1$, $z_{\text{mar},i}$ is given in Eq.~\eqref{zmarginal} and $z_{\text{mob},i}(R_e)=1+\f{L_s}{\sqrt{3}L_s^\ast}-\f{2\sqrt{\k^2L^2_s-E_s^2}}
  {\sqrt{3}\k L_s^\ast}$. The value of $L_0$ is obtained for $z_{\text{mob},o}=0$.
Meanwhile, we display in Fig.~\ref{fig:asy2} the corresponding values of $z_{\text{mob}}$ and $\M{P}_e$, in which $E_s$ is chosen to be $0$ for the bounded, the plunging and the anti-plunging emitters, and $E_s$ is chosen to be $-\k L_s$ for the deflecting emitters.
In order to see the effects of $E_s$, we display $z_{\text{mob}}(R_s)$ and $\M{P}_e(R_s)$ for the emitters with $L_s=\sqrt{3}L_s^\ast$ and $\f{4}{5}L_s^\ast$, and each with several values of $E_s$ in  Fig.~\ref{fig:asy3}. The effects of $E_s$ can also be seen from Fig.~\ref{fig:escapetogether}.
Fig.~\ref{fig:asy2} shows that, if $E_s$ is fixed,  $z_{\text{mob}}(L_s)$ and $\M{P}_e(L_s)$ for the near-NHEK emitters have similar behaviors as those for the NHEK emitters. However, the values of PEP and MOB do depend on $E_s$ and tend to the NHEK values as $\ab{E_s}\rightarrow\infty$.
For $L_s\rightarrow L_0=\f{L_s^\ast}{\sqrt{5}}$ and $E_s=0$, the turning point $R_+$ approaches $\td R_1$ and the maximum PEP is about $31\%$.
For $L_s\geq L_s^\ast$ and $E_s\rightarrow E_c=-\k\sqrt{-C_s}$, the PEP for photon emissions from unstable circular orbits at $R_c=\k L_s/\sqrt{-C_s}$ can be formally written as $\M{P}_{\text{cir},u}=\M{P}_e(L_s)$.
We find that $\M{P}_{\text{cir},u}$ is the same as the PEP of the marginal infalling emitters in the NHEK region, $\M{P}_{\text{mar},i}$, that is $\M{P}_{\text{cir},u}=\M{P}_{\text{mar},i}$ (see the dotted blue curve on the right in Fig.~\ref{fig:asy2} and the lightblue curve above the lightgreen region on the right in Fig.~\ref{fig:asy1}).
As $E_s$ is perturbed from $E_c$ for $L_s\in[L_s^\ast,\sqrt{3}L_s^\ast)$, an unstable circular orbits would become a deflecting/anti-plunging/plunging orbit and $\M{P}_e(L_s)$ starts to depart from $\M{P}_{\text{cir},u}$.
Among all the plunging emitters with a given $L_s\in[L_s^\ast,\sqrt{3}L_s^\ast)$, the PEP takes a maximum value for $E_s\rightarrow E_c^+$, which corresponds to the marginal plunging emitter starting from $R_c$.
For the marginal plunging emitter with $L_s=L_s^\ast$, the orbital radius $R_c$ tends to the radius of the ISCO, and we have $E_c=0$ and
\be
\M{P}_e(R_{\text{ISCO}})|_{(L_s^\ast,0,-1)}\approx55\%,\qquad
\M{P}_e(R_z)|_{(L_s^\ast,0,-1)}\approx12\%,
\ee
where
\be
R_z=\sqrt{1+\f{2}{\sqrt{3}}}\k.
\ee
Here we only give a few selected illustrations for several representative features for photon emissions from the near-horizon emitters and a complete picture can be built up from the figures in this paper.

\section{Summary and conclusion}\label{sec:summary}

In this paper, we studied isotropic and monochromatic photon emissions from equatorial emitters moving on timelike geodesics in the (near-)NHEK regions [see Eqs.~\eqref{NKmetric} and \eqref{nNKmetric}]. The orbits for these geodesics were classified in Tables \ref{table:NKorbits} and \ref{table:nNKorbits} whose orbital radii were written as
$r_s=M(1+\e^q R_s)$.
Along each emitter's orbit, we had the conserved quantities for the emitter: the angular momentum $l_s$ and the energy $\o_s$. For a (near-)NHEK emitter, we introduced
$l_s=L_s$ and
$\o_s=\f{l_s}{2M}+\f{E_s}{2M}\e^q $.
We had $q=\f{2}{3}$ for the NHEK case in the ISCO scale and $q=1$ for the near-NHEK case. In Sec.~\ref{sec:NKemitters}
we computed the PEP and MOB for photon emissions from these emitters.
In general, the results depend on the emitter radius $R_s$ and the constants of the emitter's motion: $L_s$, $E_s$ and $s_r$, with $s_r$ being the radial orientation.
Then, in Sec.~\ref{sec:dependenceonR} we analyzed the behaviors of PEP and MOB along each given orbit to see their dependence on $R_s$, and we summarized the results in Tables \ref{table:NKbehavior2} and \ref{table:nNKbehavior2}. After that, we analyzed the effects of an emitter's proper motion on the MOB and PEP in Sec.~\ref{motioneffects}.
The main analytical result for MOB was summarized in Eqs.~\eqref{zmobout}, \eqref{zmobin} and \eqref{zmobinside}. Moreover, the numerical results for MOB and PEP were displayed in Figs.~\ref{fig:escapeandmob}, \ref{fig:escapetogether}, \ref{fig:asy1}, \ref{fig:asy2} and \ref{fig:asy3}, from which we found that the MOB and PEP had similar behaviors as the parameters of the emitters vary.

In Fig.~\ref{fig:escapeandmob}, we showed that the PEP and MOB for the emitters belonging to distinct classes of motions were different. In addition, in Fig.~\ref{fig:escapetogether} we displayed more examples on the PEP for the emitters belonging to each class of motion. Furthermore, from Figs. \ref{fig:asy1}, \ref{fig:asy2} and \ref{fig:asy3} we could see that the overall observability of an emitter, that said the magnitudes of PEP and MOB, were determined by the angular momentum $l_s$ of the emitter. This is  reasonable since the energy $\o_s$ of a (near-)NHEK emitter is constrained to be near the superradiant bound $\o_\ast=\f{l_s}{2M}$. And, as expected, the photons from the outgoing emitters are brighter than those from the infalling ones that have the same parameters $(l_s,\o_s)$.
As the angular momentum $l_s$ of an emitter increases, the MOB and the maximum of PEP for the  emitters with all allowed motions increase accordingly.
On the other hand, for the emitters with the same angular momentum $l_s$, the corrections of their energies to the critical energy,\footnote{Note that, since in the NHEK region we have $\D \o$ at $\M{O}(\e^{2/3})$ while in the near-NHEK region we have $\D \o$ at $\M{O}(\e)$,  the near-NHEK emitters correspond to the NHEK emitters with $\D\o=0$.} $\D \o=\o_s-\o_\ast=\f{E_s}{2M}\e^q$, distinguish their motions and affect the MOB and PEP for these emitters as well. Simply speaking, while $l_s$ plays a dominant role on the overall feature of MOB and PEP, $\D \o$ starts to play a role after $l_s$ being fixed.

We picked out several characteristic values of $l_s$ to illustrated our results.
If $l_s=0$, then $z_{\text{mob},o}=0$ [see Eqs.~\eqref{zmobout} and \eqref{zmobinside}] which means that no photon emissions from retrograde emitters ($l_s<0$) were blueshifted. We focused on the case $l_s>0$, which has blueshifted photon emissions.
Around $l_s=l_s^\ast$ [see Eq.~\eqref{lsstar}], the emitters start to appear with distinct motions and different observational features. For $0<l_s<l_s^\ast$, the emitters are bounded and require $\D \o<-\e^q \k \o_\ast $, and the PEP for them are always less than $59\%$. In this case,  the maximum value is reached  as $l_s\rightarrow l_s^\ast$ and the emitter is bounced outward at the inner boundary of its orbit. For $l_s=l_s^\ast$ and $\D \o=0$,  the emitters are on the ISCO in the NHEK region and have the marginal (anti-)plunging motion in the near-NHEK region, while for $l_s=l_s^\ast$ and $\D \o\gtrless0$, the emitters have the (anti-)plunging/deflecting motions, respectively. And, for the emitter at the ISCO, the PEP is about $55\%$ and the MOB is $1-1/\sqrt{3}\approx0.42$.
Similarly, for $l_s>l_s^\ast$, the emitters with $\D\o=-\e \k\o_\ast\sqrt{-C_s/l_s^2}$ are on unstable circular orbits while the emitters with $\D \o\gtrless-\e^q \k\o_\ast\sqrt{-C_s/l_s^2}$ take the plunging (or anti-plunging)/deflecting motions, respectively.
The PEP for the emitter on unstable circular orbits, $\M{P}_{\text{cir},u}$, decreases from $55\%$ to $50\%$ as the orbital radius decreases from the ISCO radius towards the horizon one.
The PEP for the deflecting and anti-plunging emitters are always greater than $50\%$ and the MOB for the outgoing parts of the orbits are always greater than $0.42$,  which implies that the emitters with $l_s>l_s^\ast$ and with outward radial motions could be well observed by distant observers. On the other hand,
the PEPs for the plunging emitters decreases as the emitters move from the NHEK boundary towards the horizon and are always less than $55\%$ (the value for the emitters at the ISCO). We found that $\M{P}_{\text{mar},i}=\M{P}_{\text{cir},u}$ with $\M{P}_{\text{mar},i}$ the PEP for the plunging emitters at the NHEK boundary. Moreover, as a plunging emitter moves from NHEK infinity towards the horizon, the inward radial velocity for the emitter with $\D\o>0$ increases monotonously while that for the emitter with $\D\o<0$ decreases at the beginning and then increases to the end.
The maximum value of MOB for a plunging emitter with $\D\o>0$ is obtained at the NHEK infinity, and this maximum value ($z_{\text{mar},i}$) decreases as $l_s$ is increased and $z_{\text{mar},i}=0$ for $l_s=\sqrt{3}l_s^\ast$. In addition,  as $l_s$ increases for $\D\o\rightarrow-\e \k\o_\ast\sqrt{-C_s/l_s^2}$, the maximum value of MOB for a plunging emitter increases, and this maximum value ($z_{\text{cir},u}$) is obtained at the corresponding orbital radius of an unstable circular orbit and actually $z_{\text{cir},u}=z_{\text{mob},o}$.
Furthermore, we found that the PEP and MOB decreased faster for the plunging emitters with larger $\D\o$, which means that for the plunging emitters, a smaller $\D\o$ would extend the observable range towards the deeper region in the near-horizon throat of a high-spin black hole.

\section*{Acknowledgments}
We thank Peng-Cheng Li for his helpful discussions and comments on the manuscript. The work is in part supported by NSFC Grant  No. 11735001.

\bibliographystyle{utphys}
\bibliography{note}

\end{document}